%% file: order_by_order_paper.tex
\documentclass[aps,prc,reprint,showpacs,superscriptaddress,onecolumn,notitlepage]{revtex4-1}

\usepackage{graphicx,color} 
\usepackage{bm}       
\usepackage{amsmath} 
\usepackage[dvipsnames]{xcolor}
\usepackage{ulem}
\usepackage{siunitx}
\usepackage{multirow}
\usepackage{gensymb}

\begin{document}

\newcommand {\nc} {\newcommand}

\newcommand{\vv}[1]{{$\bf {#1}$}}
\newcommand{\vvm}[1]{{\bf {#1}}}
\def\btau{\mbox{\boldmath$\tau$}}

\nc {\IR} [1]{\textcolor{red}{#1}}
\nc {\IB} [1]{\textcolor{blue}{#1}}
\nc {\IP} [1]{\textcolor{magenta}{#1}}
\nc {\IM} [1]{\textcolor{Bittersweet}{#1}}
\nc {\IE} [1]{\textcolor{Plum}{#1}}

\nc{\ninej}[9]{\left\{\begin{array}{ccc} #1 & #2 & #3 \\ #4 & #5 & #6 \\ #7 & #8 & #9 \\ \end{array}\right\}}
\nc{\sixj}[6]{\left\{\begin{array}{ccc} #1 & #2 & #3 \\ #4 & #5 & #6 \\ \end{array}\right\}}
\nc{\threej}[6]{ \left( \begin{array}{ccc} #1 & #2 & #3 \\ #4 & #5 & #6 \\ \end{array} \right) }
\nc{\half}{\frac{1}{2}}
\nc{\numberthis}{\addtocounter{equation}{1}\tag{\theequation}}
\nc{\lla}{\left\langle}
\nc{\rra}{\right\rangle}
\nc{\lrme}{\left|\left|}
\nc{\rrme}{\right|\right|}

\title{\textit{Ab initio} nucleon-nucleus elastic scattering with chiral effective field theory uncertainties}

\author{R.~B.~Baker}
\affiliation{Institute of Nuclear and Particle Physics, and Department of Physics and Astronomy, Ohio University, Athens, OH 45701, USA}

\author{B.~McClung}
\affiliation{Institute of Nuclear and Particle Physics, and Department of Physics and Astronomy, Ohio University, Athens, OH 45701, USA}

\author{Ch.~Elster}
\affiliation{Institute of Nuclear and Particle Physics, and Department of Physics and Astronomy, Ohio University, Athens, OH 45701, USA}

\author{P.~Maris}
\affiliation{Department of Physics and Astronomy, Iowa State University, Ames, IA 50011, USA}

\author{S.~P.~Weppner}
\affiliation{Natural Sciences, Eckerd College, St. Petersburg, FL 33711, USA}

\author{M.~Burrows}
\affiliation{Department of Physics and Astronomy, Louisiana State University, Baton Rouge, LA 70803, USA}

\author{G.~Popa}
\affiliation{Institute of Nuclear and Particle Physics, and Department of Physics and Astronomy, Ohio University, Athens, OH 45701, USA}

\date{December 4, 2021}

\begin{abstract}
\begin{description}
\item[Background] 
Effective interactions for nucleon-nucleus elastic scattering from first principles require the use of
the same nucleon-nucleon interaction in the structure and reaction calculations, and a
consistent treatment of the relevant operators at each order.

\item[Purpose] 
Systematic investigations of the effect of truncation uncertainties of chiral nucleon-nucleon ($NN$) forces
have been carried out for scattering observables in the two- and three-nucleons system as well as 
for bound state
properties of light nuclei. Here we extend this type of study to proton and neutron
elastic scattering for $^{16}$O and $^{12}$C. 

\item[Methods] 
Using the frameworks of the spectator expansion of multiple scattering theory as well as the no-core shell
model, we employ one specific chiral interaction from the LENPIC collaboration and consistently
calculate the leading order effective nucleon-nucleus interaction up to the third chiral order
(N2LO), from which we extract elastic scattering observables.
 Then we apply pointwise as well as correlated uncertainty quantification
for the estimation of the chiral truncation error.

\item[Results] 
We calculate and analyze proton elastic scattering observables for $^{16}$O and neutron elastic
scattering observables for $^{12}$C between 65 and 185~MeV projectile kinetic energy. We find 
qualitatively similar results for the chiral truncation uncertainties as in few-body systems, and assess them using 
similar diagnostic tools. The order-by-order convergence of the scattering observables for $^{16}$O 
and $^{12}$C is very reasonable around 100~MeV, while for higher energies the chiral 
expansion parameter becomes too large for convergence. We also find a nearly perfect correlation between the
differential cross section for neutron scattering and the $NN$ Wolfenstein amplitudes for small
momentum transfers.

\item[Conclusions]
The diagnostic tools for studying order-by-order convergence of a chiral $NN$ interaction in
observables in few-body systems can be employed for observables in nucleon-nucleus scattering with only
minor modifications provided the momentum scale in the problems is not too large. 
We also find that the chiral $NN$ interaction on which our study is based on, gives a very good
description of differential cross sections as well as spin observables for $^{16}$O and $^{12}$C as low
as 65~MeV projectile energy. In addition, the very forward direction of the neutron differential cross
section mirrors the behavior of the $NN$ interaction amazingly well.

\end{description}
\end{abstract}


\maketitle

\section{Introduction}
\label{sec:intro}

\input{intro.tex}


\section{Theoretical Frameworks}
\label{sec:theory}

\input{formal.tex}


\section{Results and Discussion}
\label{sec:results}

\input{results.tex}


\section{Conclusions and Outlook}
\label{sec:conclusions}

\input{conclusions.tex}


\begin{acknowledgments}
The authors appreciate the many useful discussions with K.~D.~Launey. R.~B.~B. and Ch.~E. gratefully acknowledge fruitful discussions with R.~J.~Furnstahl about quantifying truncation errors in EFTs. Ch.~E. acknowledges useful discussions with A.~Nogga about the LENPIC chiral $NN$ interactions.
This work was performed in part under the auspices of the U.~S.~Department of Energy under contract Nos.~DE-FG02-93ER40756 and DE-SC0018223, and by the U.~S.~NSF (PHY-1913728). The numerical computations benefited from computing resources provided by the Louisiana Optical Network Initiative and HPC resources provided by LSU, together with resources of the National Energy Research Scientific Computing Center, a U.~S.~DOE Office of Science User Facility located at Lawrence Berkeley National Laboratory, operated under contract No.~DE-AC02-05CH11231.

\end{acknowledgments}


\bibliography{denspot,clusterpot,ncsm,reactions}

\clearpage

\begin{table}
\begin{center}
\begin{tabular}{ccc|ccc}
\multicolumn{3}{c}{$^{16}$O}									&	\multicolumn{3}{c}{$^{12}$C}	\\ \hline
$E_{\mathrm{lab}}$ (MeV)		&	$p_{NA}$ (MeV)	&	$Q$		& $E_{\mathrm{lab}}$ (MeV)	&	$p_{NA}$ (MeV)	&	$Q$		\\ \hline	
65		&	333.075	&	0.55								& 65		&	326.319	&	0.54	\\
100		&	415.984	&	0.69								& 95		&	396.645	&	0.66	\\
135		&	486.598	&	0.81								& 155	&	512.013	&	0.85\\
180		&	566.646	&	0.94								&  185	&	562.242	&	0.93	\\ \hline
\end{tabular}
\end{center}
\caption{Center-of-mass momentum $p_{NA}$ and associated expansion parameter $Q$ for various energies in elastic nucleon scattering from $^{16}$O and $^{12}$C. Values for $Q$ are calculated assuming $\Lambda_b = 600$ MeV.}
\label{tab1}
\end{table}

\clearpage

\begin{figure}
\begin{center}
\begin{tabular}{c}
\includegraphics[width=0.5\textwidth]{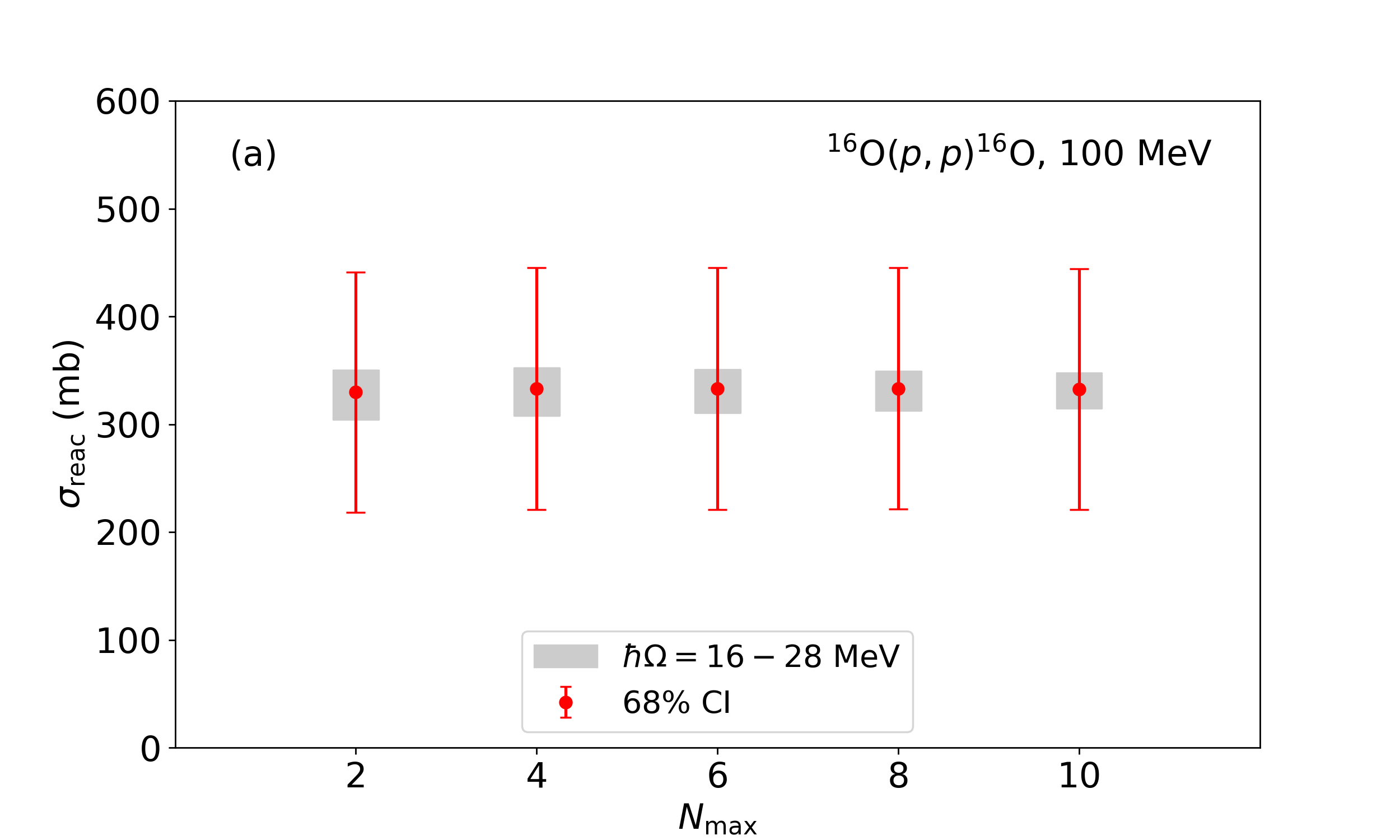}	\\
\includegraphics[width=0.5\textwidth]{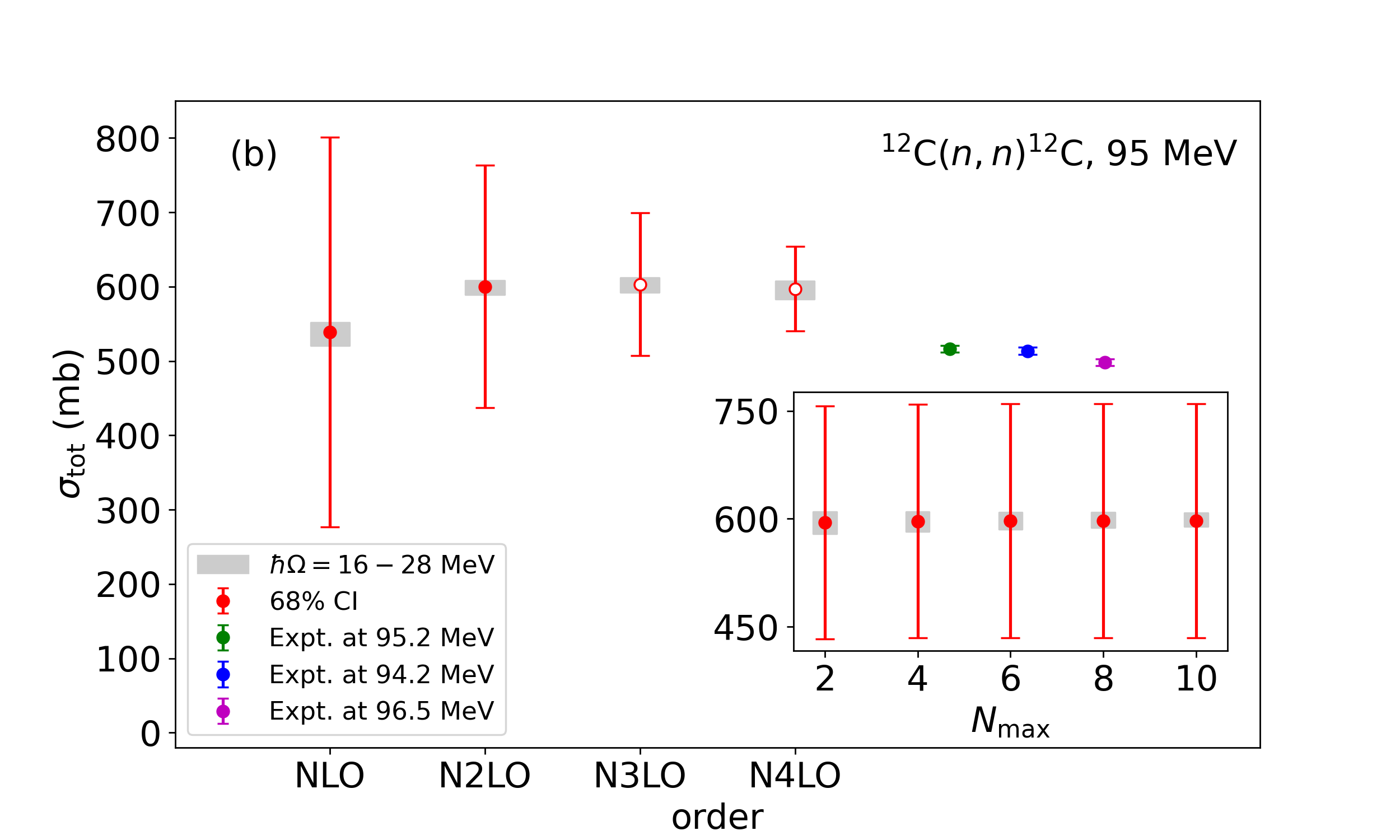}
\end{tabular}
\caption{(a) Reaction cross section for $^{16}$O$(p,p)^{16}$O for 100 MeV at N2LO as a function of $N_{\mathrm{max}}$. The gray shaded regions show variations in $\hbar\Omega$. The red error bars are the $68\%$ CIs resulting from using the pointwise approach on the LO, NLO, and N2LO results at $\hbar\Omega=20$ MeV with $Q=0.69$ and $y_{\mathrm{ref}}=$ N2LO.
(b) Total cross section for $^{12}$C$(n,n)^{12}$C for 95 MeV as a function of order compared to experimental data. The experimental value at 95.2 MeV is from Ref.~\cite{Finlay:1993hk} and the other two experimental values are from Ref.~\cite{Pruitt:2020kvc}. The inset shows the total cross section as a function of $N_{\mathrm{max}}$. See text for further discussion.
}
\label{fig1}
\end{center}
\end{figure}

\begin{figure}
\begin{center}
\includegraphics[width=1.0\textwidth]{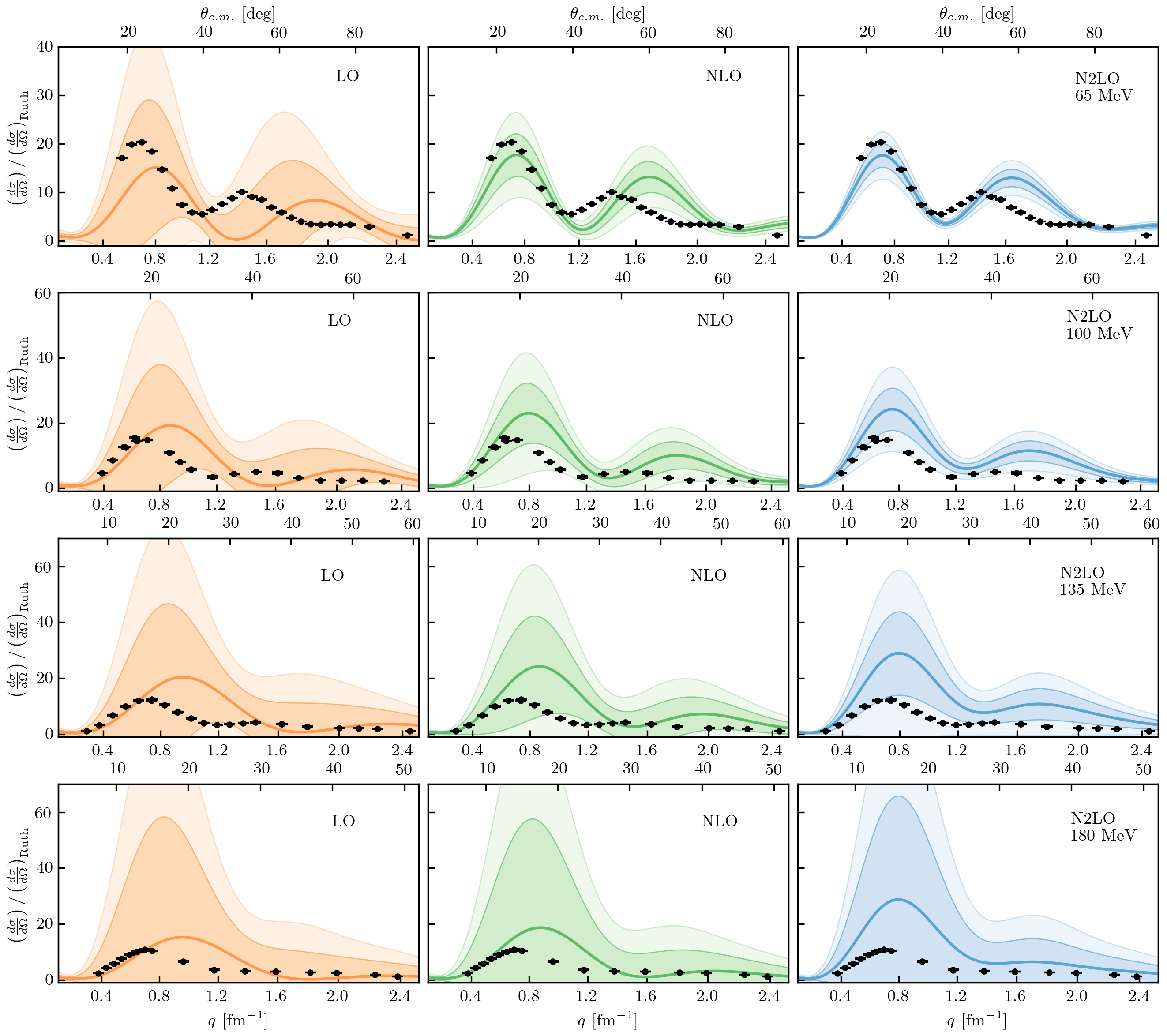}
\caption{Differential cross section divided by Rutherford at LO (left column), NLO (middle column), and N2LO (right column) with corresponding $1\sigma$ (darker bands) and $2\sigma$ (lighter bands) error bands for $^{16}$O$(p,p)^{16}$O at (first row) 65 MeV, (second row) 100 MeV, (third row) 135 MeV, and (fourth row) 180 MeV. Black dots are experimental data from Refs.~\cite{Sakaguchi:1979fpk} (65 MeV), \cite{Seifert:1990um} (100 MeV), \cite{Kelly:1989zza} (135 MeV), and \cite{Kelly:1990zza} (180 MeV). See text for further discussion.
}
\label{fig2}
\end{center}
\end{figure}

\begin{figure}
\begin{center}
\includegraphics[width=1.0\textwidth]{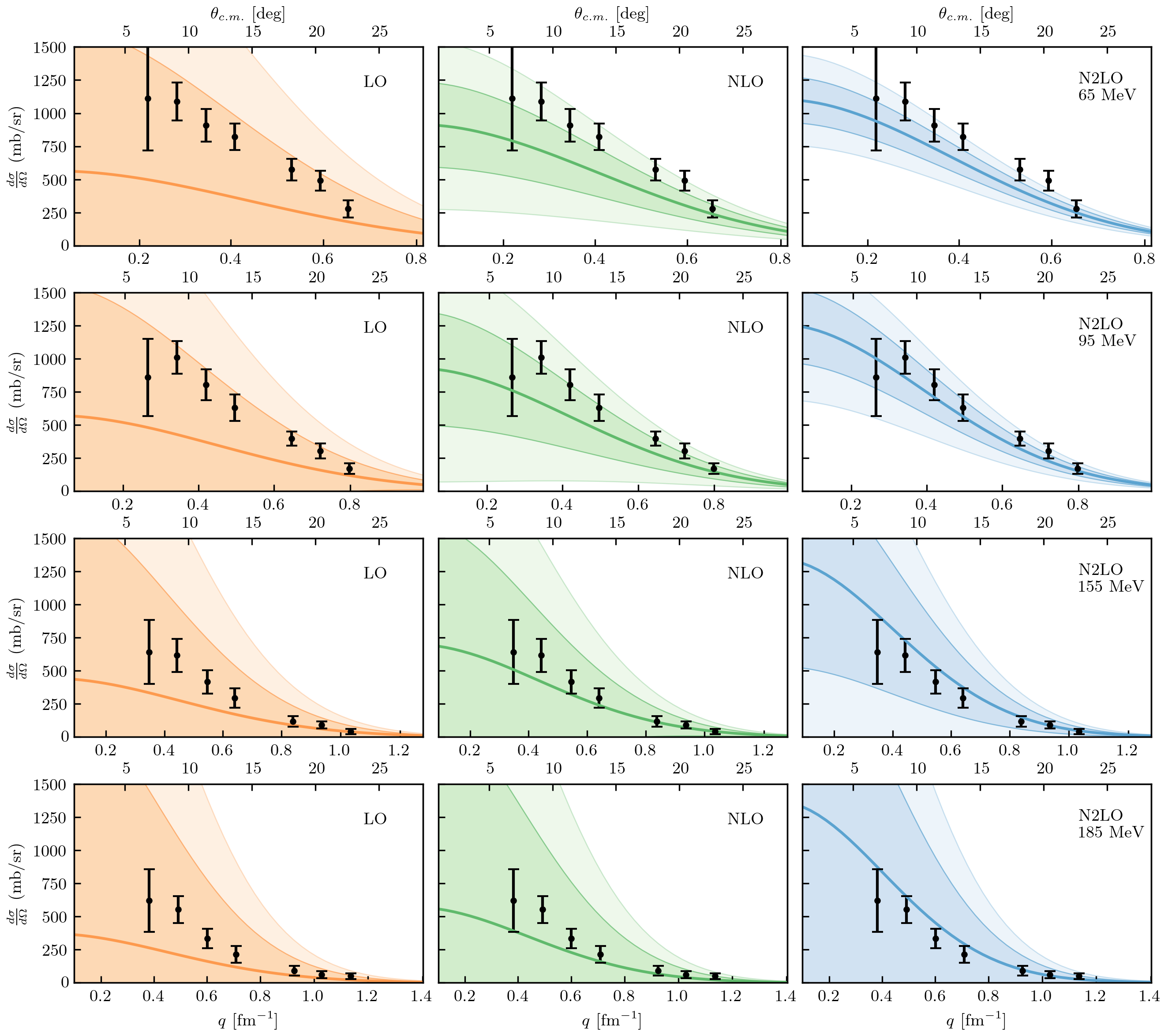}
\caption{Differential cross section at LO (left column), NLO (middle column), and N2LO (right column) with corresponding $1\sigma$ (darker bands) and $2\sigma$ (lighter bands) error bands for $^{12}$C$(n,n)^{12}$C at (first row) 65 MeV, (second row) 95 MeV, (third row) 155 MeV, and (fourth row) 185 MeV. Black dots are experimental data from Ref.~\cite{Osborne:2004vd}. See text for further discussion.
}
\label{fig3}
\end{center}
\end{figure}

\begin{figure}
\begin{center}
\begin{tabular}{cc}
\includegraphics[width=0.45\textwidth]{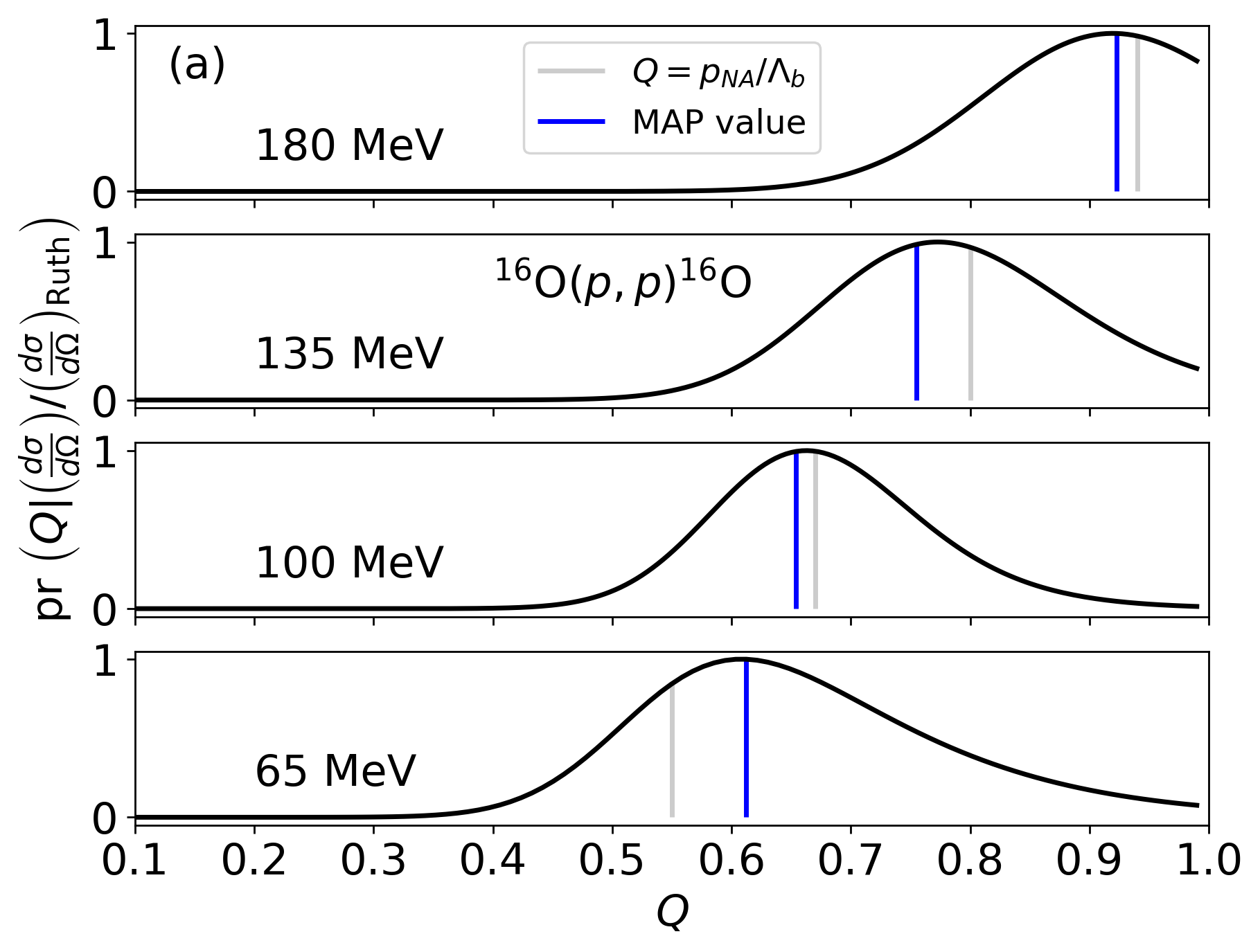}	&	\includegraphics[width=0.45\textwidth]{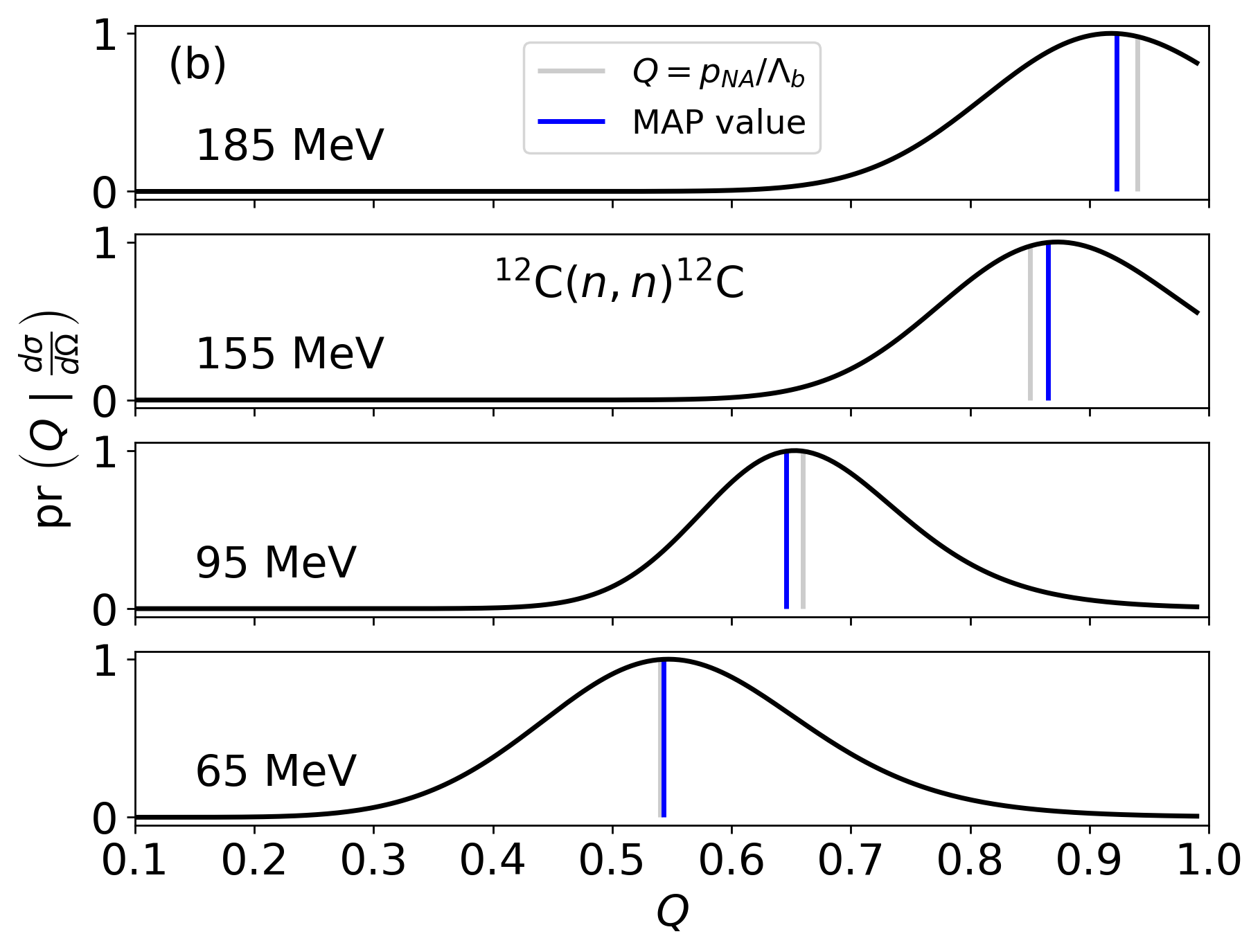}
\end{tabular}
\caption{Posterior plots for the expansion parameter $Q$ given the calculated (a) differential cross section divided by Rutherford for $^{16}$O$(p,p)^{16}$O and (b) differential cross section for $^{12}$C$(n,n)^{12}$C. The single best guess from the posteriors (MAP value) is compared to an estimate based on our choice of the relevant momentum at various energies. See text for further discussion.
}
\label{fig4}
\end{center}
\end{figure}

\begin{figure}
\begin{center}
\begin{tabular}{ccc}
\includegraphics[width=0.4\textwidth]{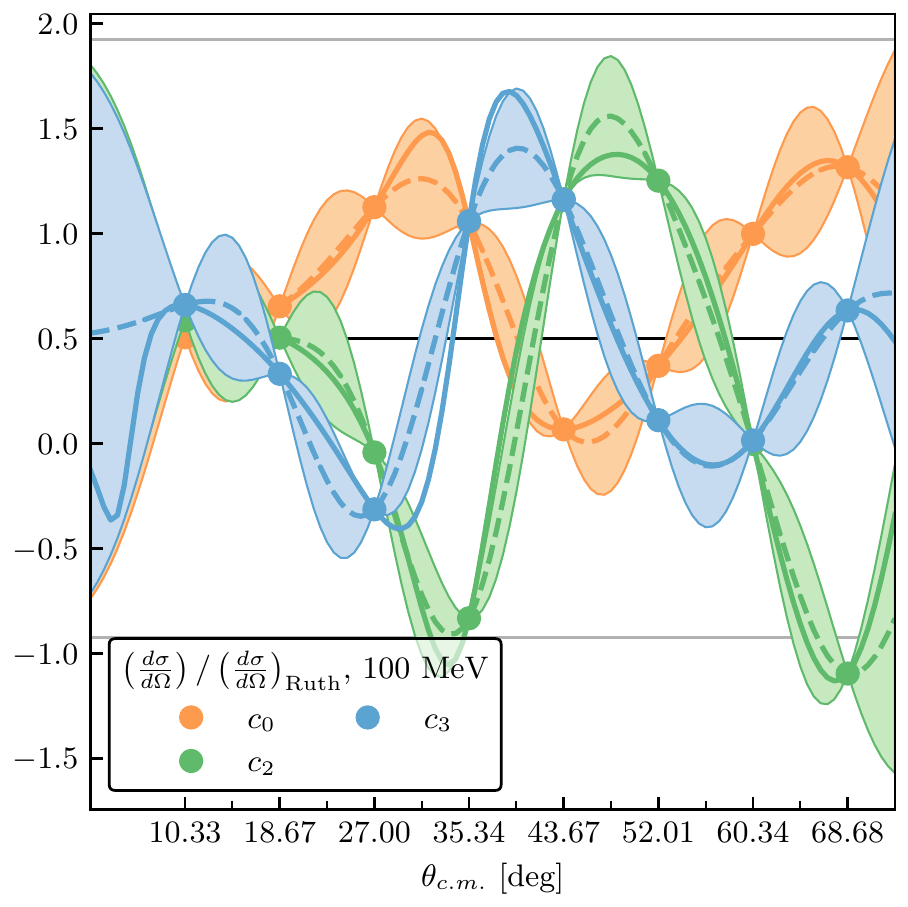}	&	\includegraphics[width=0.12\textwidth, height=0.4\textwidth]{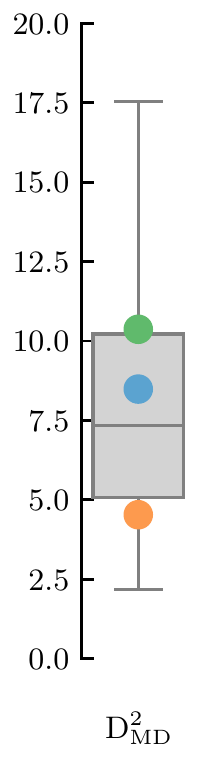} & \includegraphics[width=0.4\textwidth]{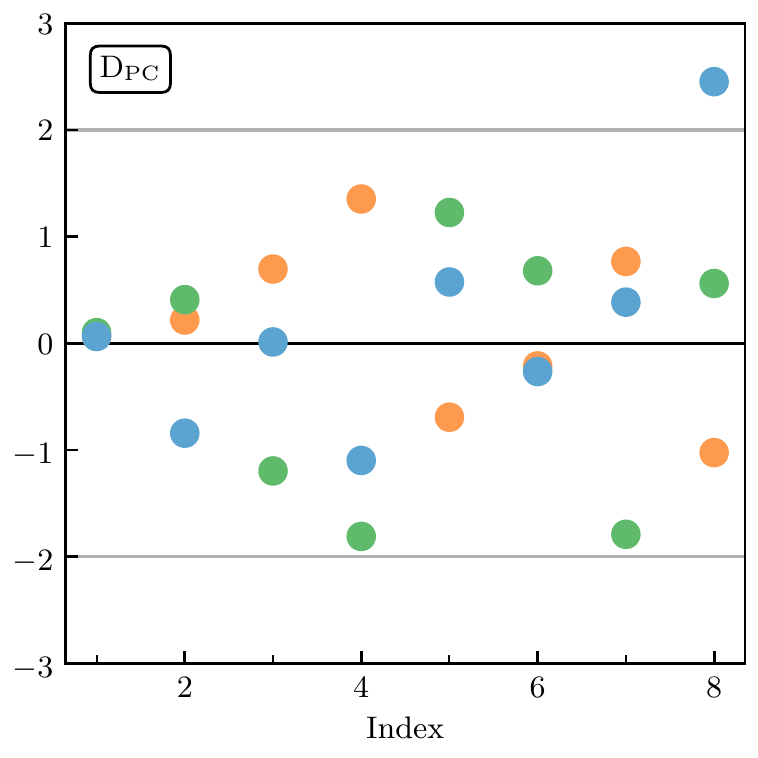}
\end{tabular}	
\caption{Coefficient curves at each order and associated diagnostics for the differential cross section (divided by Rutherford) of $^{16}$O$(p,p)^{16}$O at 100 MeV. For the coefficient curve plot, major tick marks on the $x$-axis represent training points and minor tick marks represent testing points for the Gaussian process. See text for further discussion.
}
\label{fig5}
\end{center}
\end{figure}

\begin{figure}
\begin{center}
\begin{tabular}{ccc}
\includegraphics[width=0.4\textwidth]{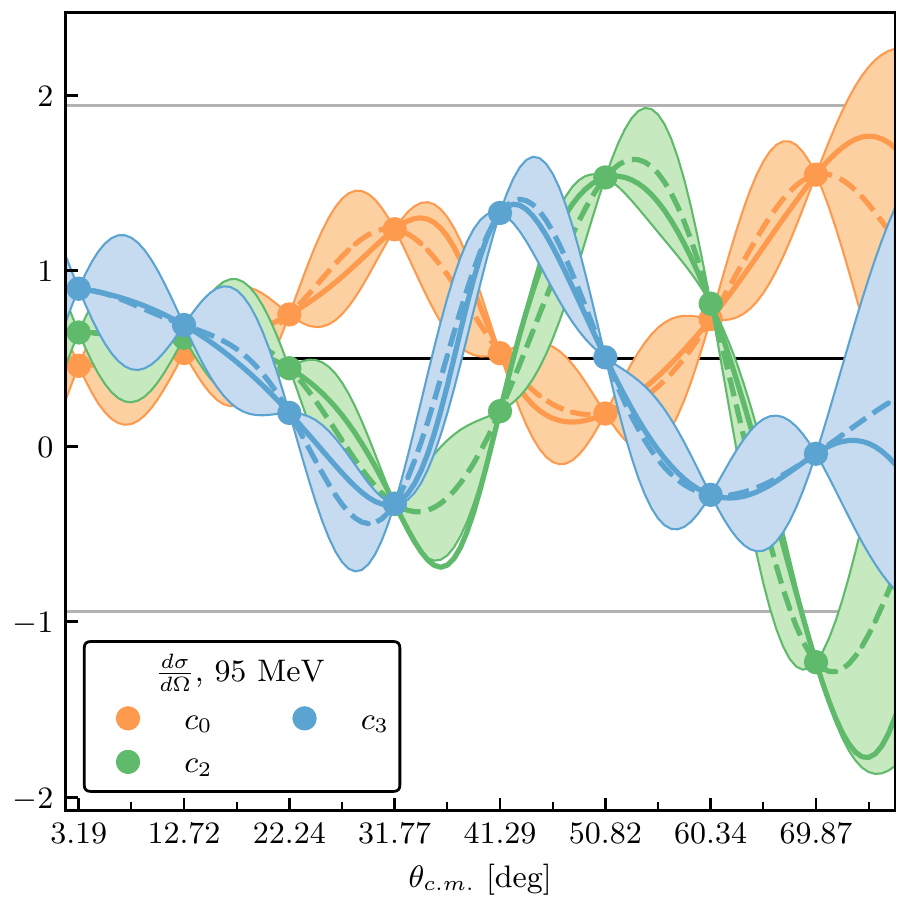}	&	\includegraphics[width=0.12\textwidth, height=0.4\textwidth]{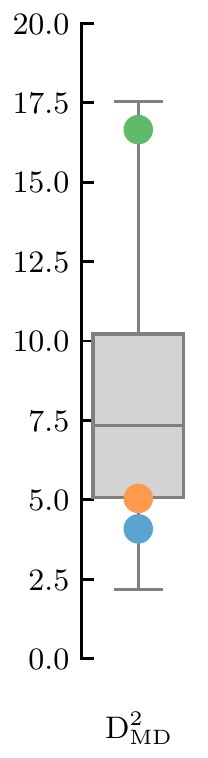} & \includegraphics[width=0.4\textwidth]{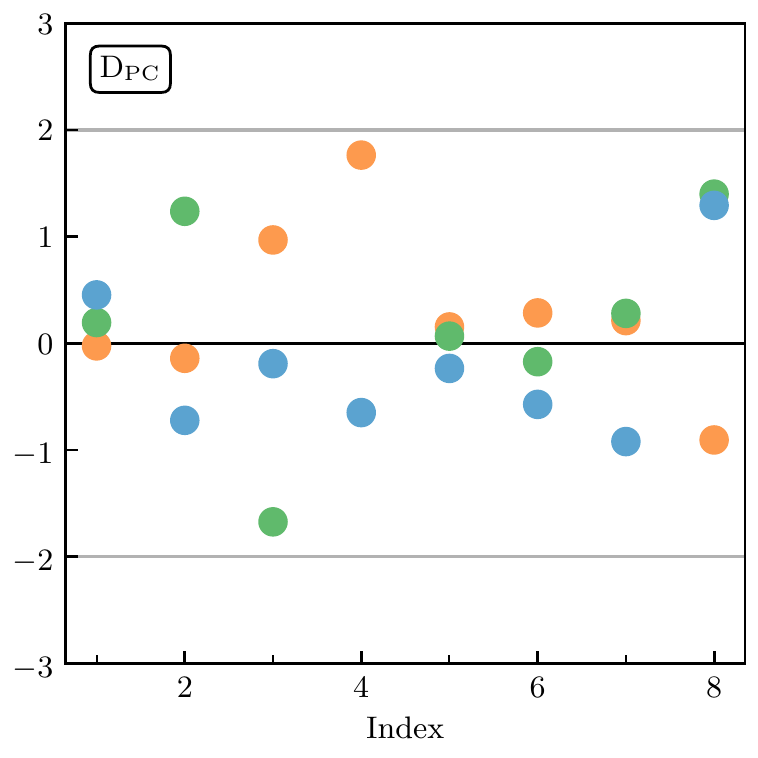}
\end{tabular}	
\caption{Coefficient curves at each order and associated diagnostics for the differential cross section of $^{12}$C$(n,n)^{12}$C at 95 MeV. For the coefficient curve plot, major tick marks on the $x$-axis represent training points and minor tick marks represent testing points for the Gaussian process. See text for further discussion.
}
\label{fig6}
\end{center}
\end{figure}

\begin{figure}
\begin{center}
\includegraphics[width=1.0\textwidth]{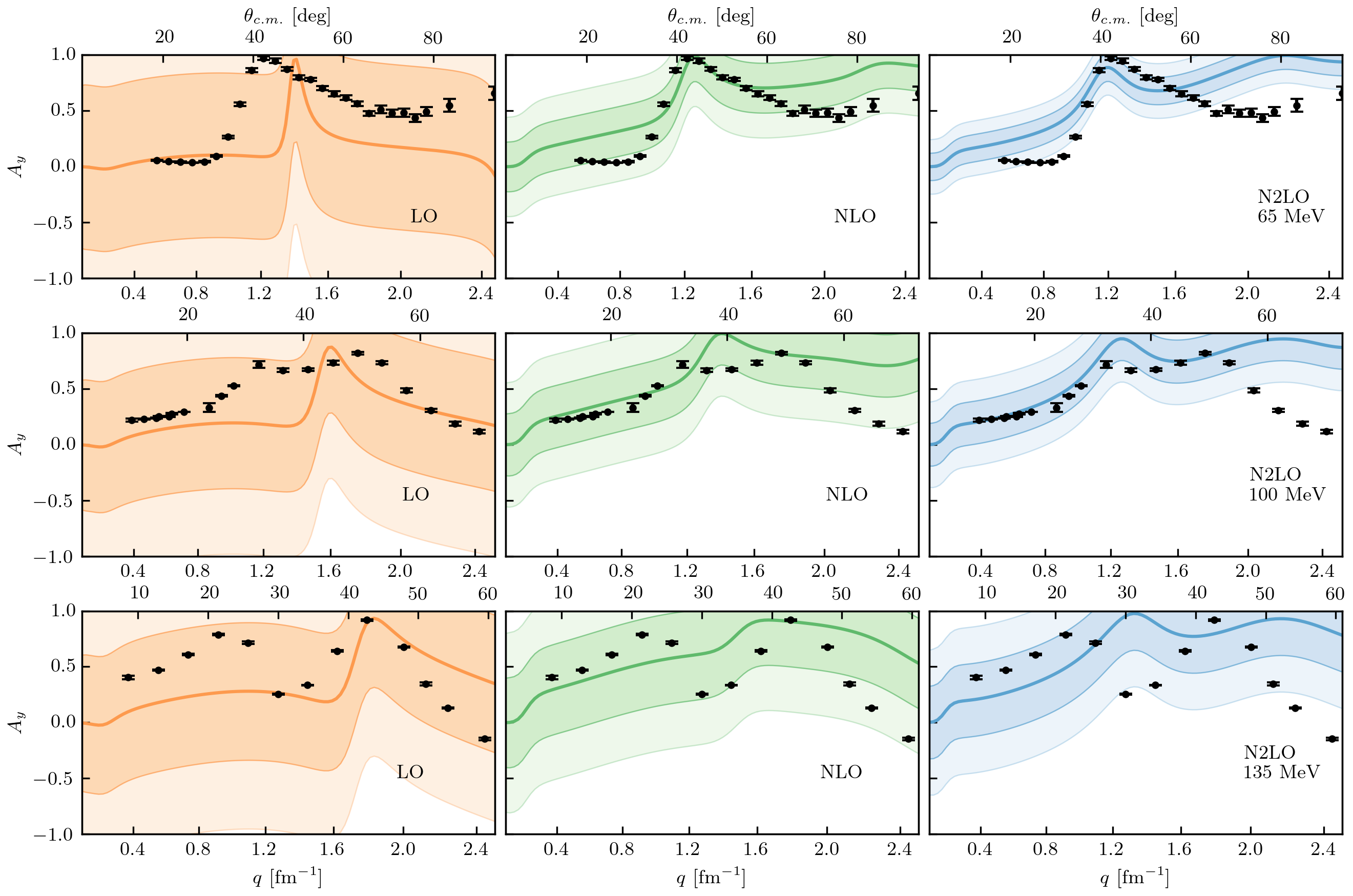} 
\caption{Analyzing power at LO (left column), NLO (middle column), and N2LO (right column) with corresponding $1\sigma$ (darker bands) and $2\sigma$ (lighter bands) error bands for $^{16}$O$(p,p)^{16}$O at (first row) 65 MeV, (second row) 100 MeV, and (third row) 135 MeV. Black dots are experimental data from Refs.~\cite{Sakaguchi:1979fpk} for 65 MeV, \cite{Seifert:1990um} for 100 MeV, and \cite{Kelly:1989zza} for 135 MeV. See text for further discussion.
}
\label{fig7}
\end{center}
\end{figure}

\begin{figure}
\begin{center}
\includegraphics[width=1.0\textwidth]{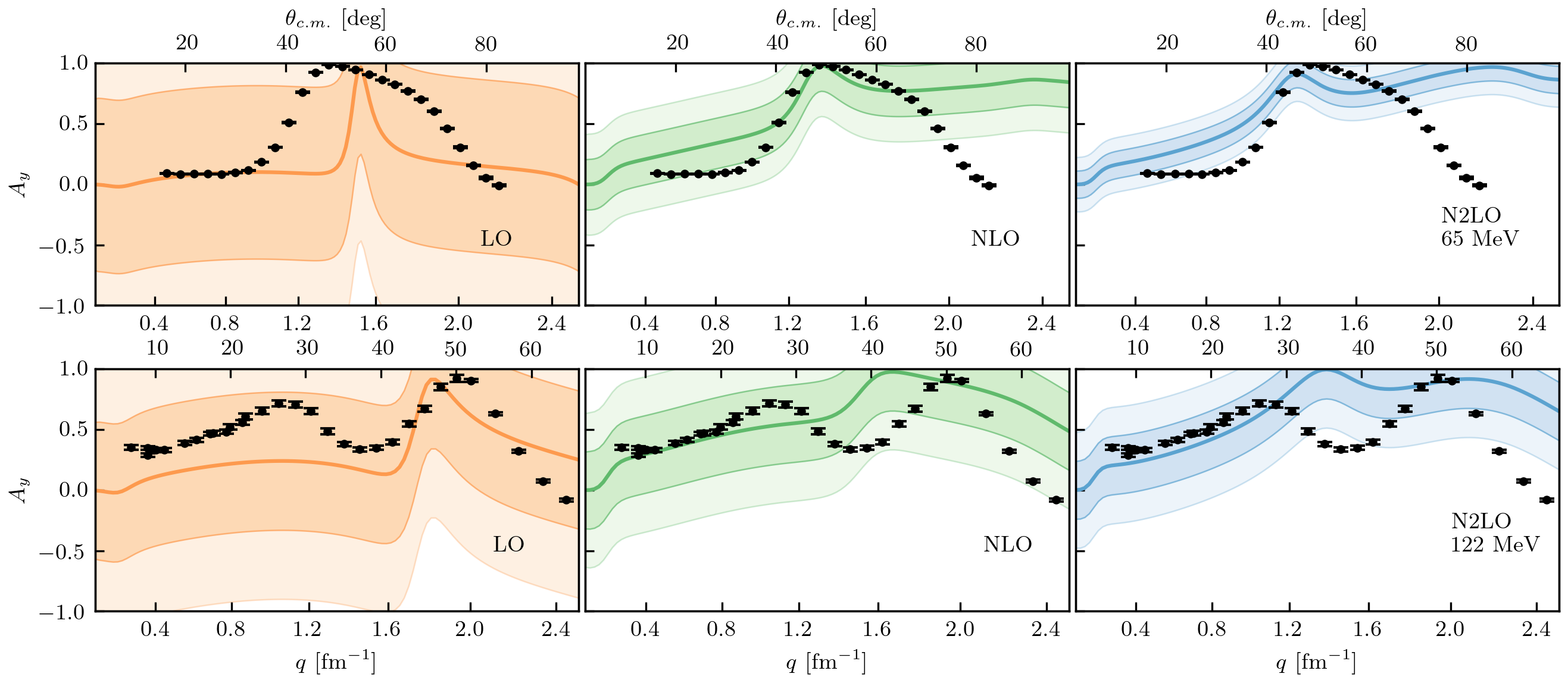}
\caption{Analyzing power at LO (left column), NLO (middle column), and N2LO (right column) with corresponding $1\sigma$ (darker bands) and $2\sigma$ (lighter bands) error bands for $^{12}$C$(p,p)^{12}$C at (first row) 65 MeV and (second row) 122 MeV. Black dots are experimental data from Refs.~\cite{Ieiri1987253} for 65 MeV and \cite{Meyer:1983kd} for 122 MeV. See text for further discussion.
}
\label{fig8}
\end{center}
\end{figure}

\begin{figure}
\begin{center}
\begin{tabular}{ccc}
\includegraphics[width=0.4\textwidth]{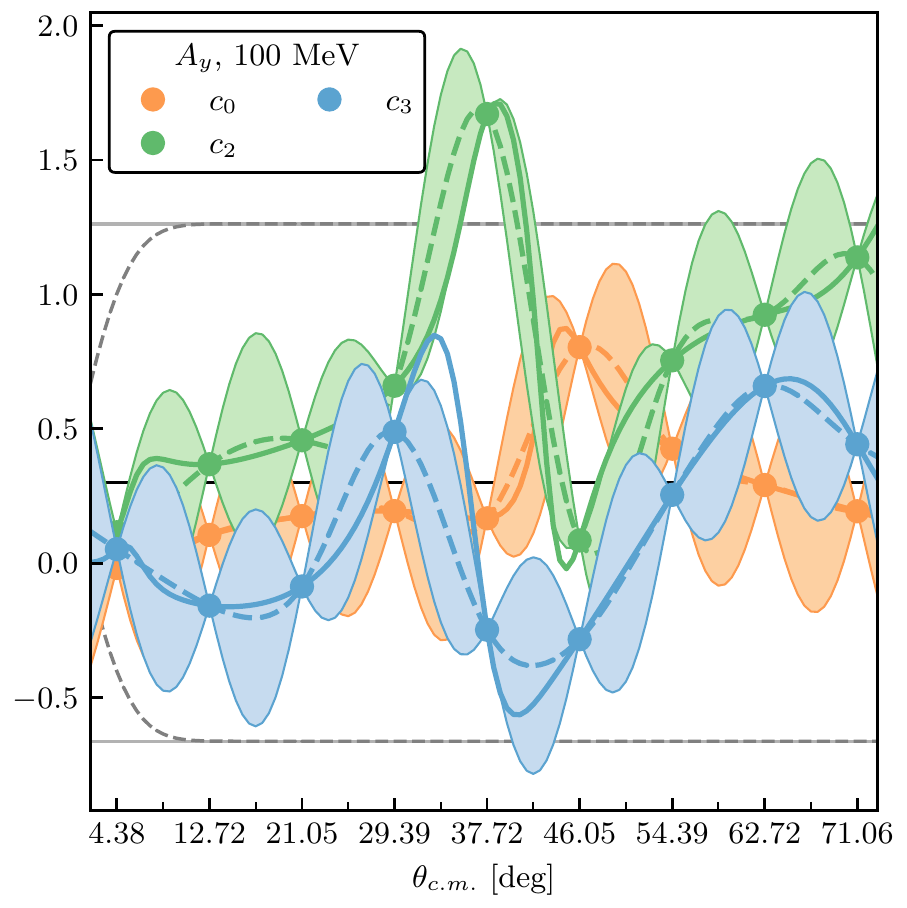}	&	\includegraphics[width=0.12\textwidth, height=0.4\textwidth]{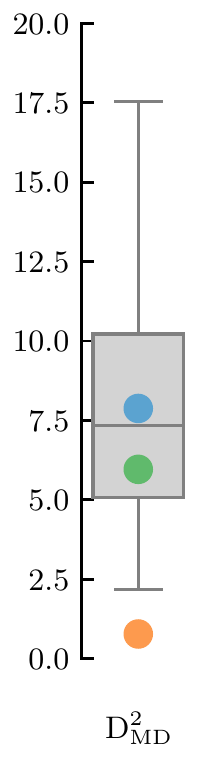} & \includegraphics[width=0.4\textwidth]{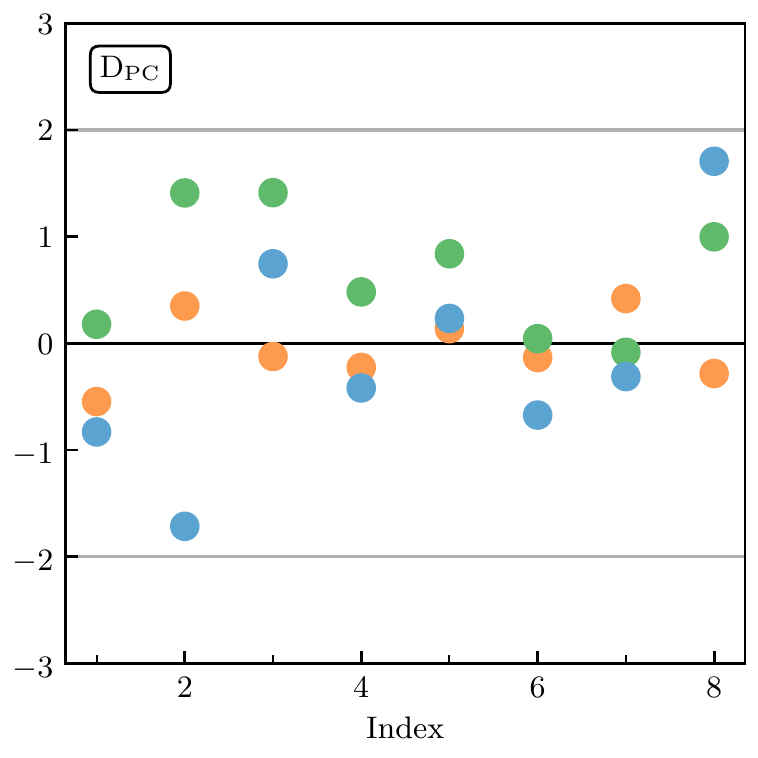} \\
\end{tabular}
\caption{Coefficient curves at each order and associated diagnostics for the analyzing power of $^{16}$O$(p,p)^{16}$O at 100 MeV. For the coefficient curve plot, major tick marks on the $x$-axis represent training points and minor tick marks represent testing points for the Gaussian process. The gray dashed line indicates the constraint on the underlying distribution which requires the $A_y$ (and the coefficients) to be 0 at $0\degree$. See text for further discussion. 
}
\label{fig9}
\end{center}
\end{figure}

\begin{figure}
\begin{center}
\includegraphics[width=1.0\textwidth]{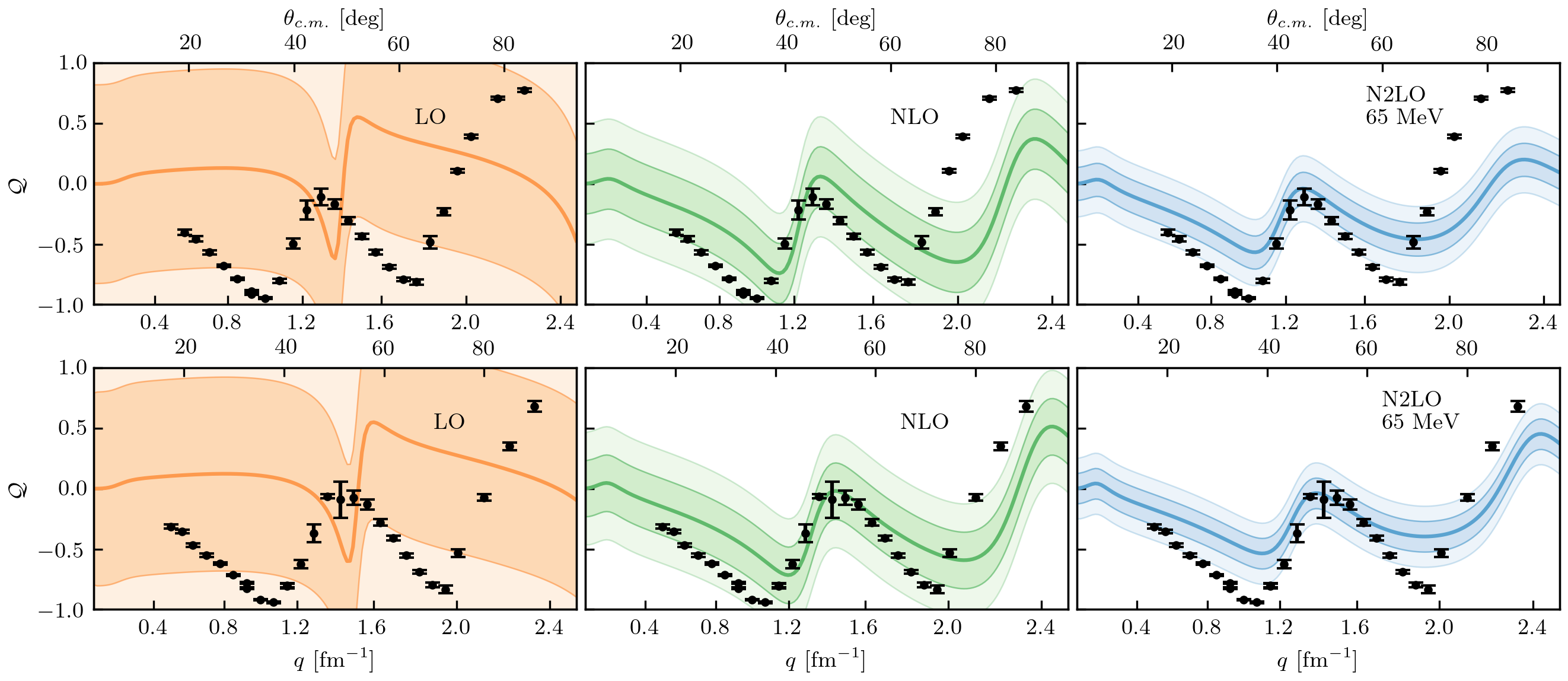} 
\caption{Spin rotation function at LO (left column), NLO (middle column), and N2LO (right column) with corresponding $1\sigma$ (darker bands) and $2\sigma$ (lighter bands) error bands for (first row) $^{16}$O$(p,p)^{16}$O and (second row) $^{12}$C$(p,p)^{12}$C, both at 65 MeV. Black dots are experimental data from Ref.~\cite{Sakaguchi:1986}. See text for further discussion.
}
\label{fig10}
\end{center}
\end{figure}

\begin{figure}
\begin{center}
\begin{tabular}{ccc}
\includegraphics[width=0.4\textwidth]{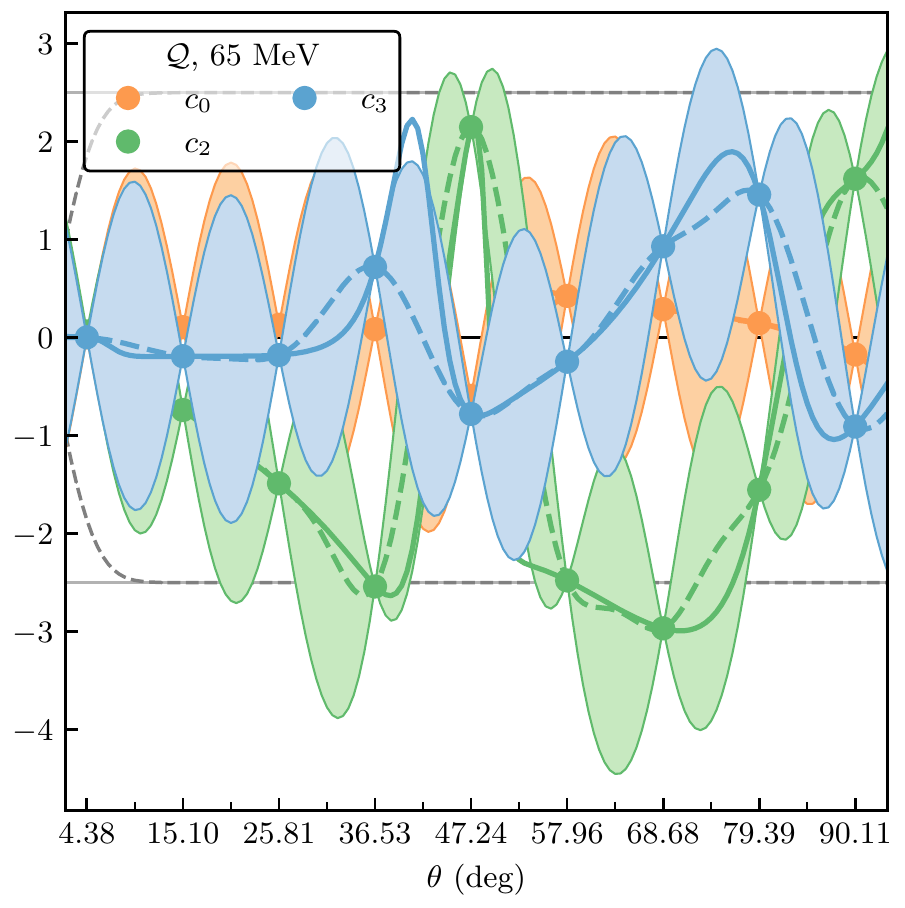}	&	\includegraphics[width=0.12\textwidth, height=0.4\textwidth]{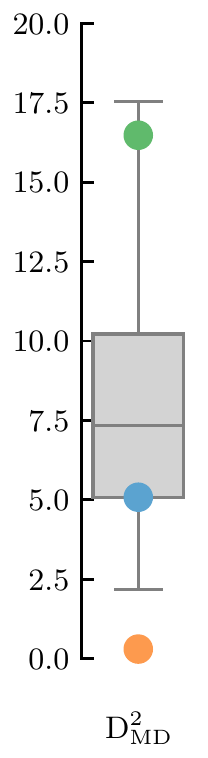} & \includegraphics[width=0.4\textwidth]{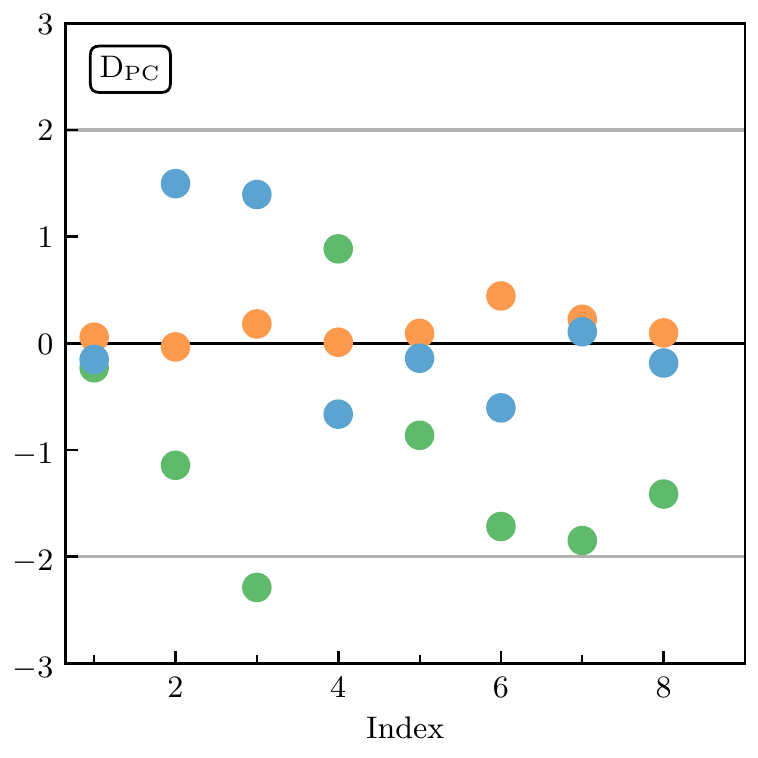} \\
\end{tabular}
\caption{Coefficient curves at each order and associated diagnostics for the spin rotation function of $^{16}$O$(p,p)^{16}$O at 65 MeV. For the coefficient curve plot, major tick marks on the $x$-axis represent training points and minor tick marks represent testing points for the Gaussian process. The gray dashed line indicates the constraint on the underlying distribution which requires the $\mathcal{Q}$ (and the coefficients) to be 0 at $0\degree$. See text for further discussion. 
}
\label{fig11}
\end{center}
\end{figure}

\begin{figure}
\begin{center}
\includegraphics[width=0.6\textwidth]{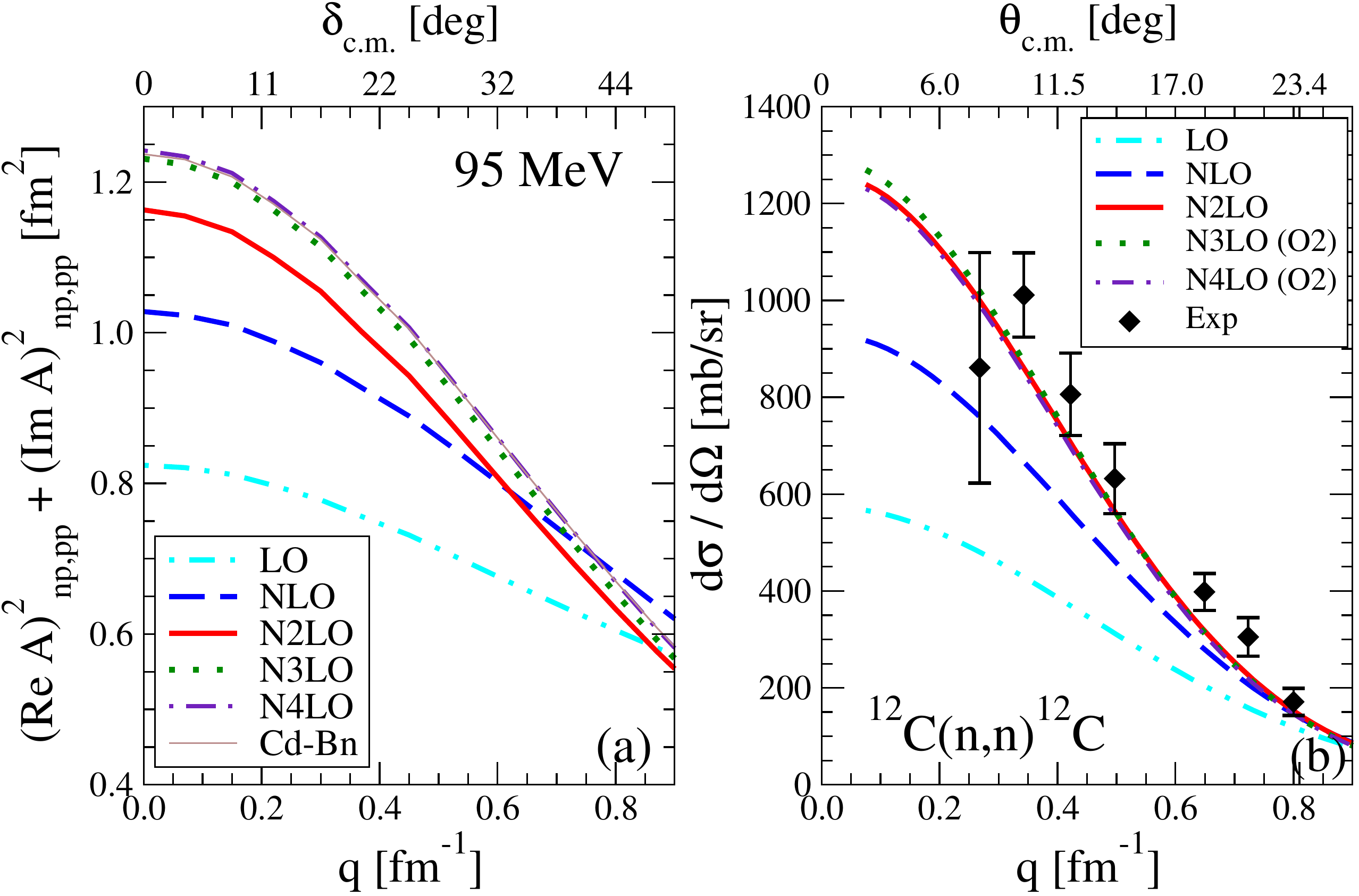}
\caption{(a) Wolfenstein amplitudes calculated at 95 MeV for various orders of 
the chiral $NN$ interaction~\cite{Epelbaum:2014efa}. The
result for the CD-Bonn potential~\cite{Machleidt:2000ge} is shown for comparison. (b) The differential cross section for $^{12}$C$(n,n)^{12}$C at 95 MeV calculated
for the same orders. The black dots and error bars show experimental data from Ref.~\cite{Osborne:2004vd}. See text for further discussion.
}
\label{fig12}
\end{center}
\end{figure}

\begin{figure}
\begin{center}
\includegraphics[width=0.45\textwidth]{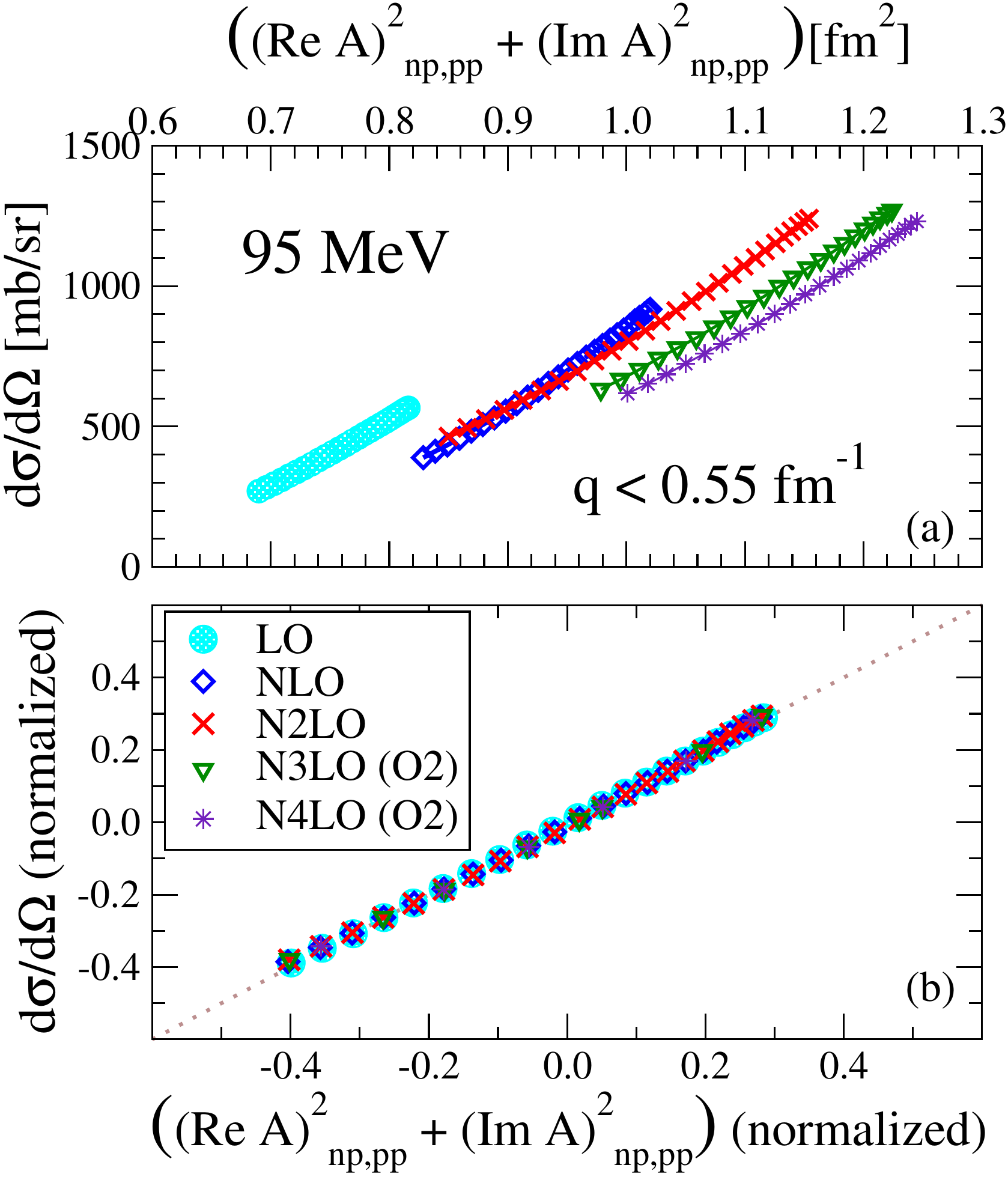}
\caption{(a) Differential cross section for $^{12}$C$(n,n)^{12}$C as a function of Wolfenstein amplitudes, both at 95 MeV for
the same orders as in Fig.~\ref{fig12}, and for values of the momentum transfer less than $0.55$ fm$^{-1}$. (b) Correlation plot of the same quantities as in (a), but normalized to illustrate the near perfect correlation, with a correlation coefficient of $0.99$. The dotted gray line indicates a perfect positive correlation. See text for further discussion.
}
\label{fig13}
\end{center}
\end{figure}

\end{document}

%% file: intro.tex

Elastic scattering of protons or neutrons from stable nuclei has traditionally played an important role
in determining phenomenological optical models or testing accuracy and validity of microscopic models
thereof. Major progress has been made in the development of nucleon-nucleon ($NN$) and three-nucleon
($3N$) interactions from chiral effective field
theory (see e.g.~\cite{Epelbaum:2014sza,Epelbaum:2014efa,Reinert:2017usi,Epelbaum:2019kcf,Machleidt:2011zz,Entem:2017gor}).
These, together with the utilization of massively parallel computing resources (e.g.,
see~\cite{LangrDDLT19,LangrDDT18,SHAO20181,CPE:CPE3129,Jung:2013:EFO}), have placed {\it ab initio}
large-scale simulations at the frontier of nuclear structure and reaction explorations. Among other
successful many-body theories, the {\it ab initio} no-core shell-model (NCSM) approach (see, e.g.,
\cite{Navratil:2000ww,Roth:2007sv,BarrettNV13,Binder:2018pgl}), has over
the last decade taken center stage in the development of microscopic tools for studying the
structure of atomic nuclei. The NCSM
concept combined with a symmetry-adapted (SA) basis in the {\it ab initio} SA-NCSM
\cite{LauneyDD16} has further expanded the reach to the structure of intermediate-mass
nuclei~\cite{Dytrych:2020vkl}. One path of extending the reach of this successful approach to describing
reactions is  the construction of {\it ab initio} effective interactions for
e.g. elastic scattering of nucleon from nuclei in the framework based on the spectator expansion~\cite{Siciliano:1977zz} of
multiple scattering theory. Recently the leading order term in the spectator expansion has been successfully
derived and calculated, treating the $NN$ interaction consistently when deriving the effective
interaction~\cite{Burrows:2020qvu,Baker:2021izp}. 

The recently developed consistent {\it ab initio} leading order nucleon-nucleus ($NA$) effective interaction 
allows the study of truncation uncertainties of the chiral $NN$ interaction in elastic $NA$ scattering
observables. In this work we focus on one specific chiral $NN$
interaction~\cite{Epelbaum:2014sza,Epelbaum:2014efa} and explore truncation uncertainties for elastic proton
scattering from $^{16}$O and  elastic neutron scattering from $^{12}$C in the energy regime from 65 to
185~MeV projectile energy. 

The choice of considering elastic scattering from these two nuclei has several motivations. 
First, we want to study nuclei 
in which the 0$^+$ ground state has contributions in the p-shell, one of them being spherical and
traditionally considered closed shell ($^{16}$O) and one being deformed and open shell ($^{12}$C).
Second, we want to consider nuclei where  reactions are  not accessible to exact few-body methods, and 
where the structure calculations are reasonably well converged within the NCSM framework.  
In addition, experimental information for neutron elastic scattering, specifically differential cross
sections in the energy regime relevant for this study, is only available for $^{12}$C~\cite{Osborne:2004vd}
as lightest measured nucleus.

The study of truncation uncertainties in the chiral $NN$ interaction in scattering observables of the $NN$ 
system has already been  successfully carried out (see e.g.~\cite{Furnstahl:2015rha,Melendez:2017phj,Melendez:2019izc}) and
extended to the nucleon-deuteron ($Nd$) 
system~\cite{Epelbaum:2019zqc} as well as to structure observables for light
nuclei~\cite{Binder:2018pgl,Maris:2020qne}. In this work we follow the procedures developed in
Refs.~\cite{Melendez:2017phj,Melendez:2019izc} and extend them to study observables in elastic $NA$
scattering. In Section~\ref{sec:theory} we briefly give  the most important results given in that work, and
point the reader to differences to be considered when going from the $NN$ system to a $NA$ system. In
Section~\ref{sec:results} we first consider reaction and total cross sections, for which a pointwise
uncertainty quantification is best suited. Then we discuss the observables in elastic scattering which depend
on the momentum transfer or equivalently the scattering angle, and apply a correlated uncertainty
quantification. We also show that the region of low momentum transfer in the differential cross section for
neutron elastic scattering can serve as unique window on the order-by-order contributions of the chiral $NN$
interaction. We conclude in Section~\ref{sec:conclusions}.

%% file: formal.tex

The fundamental idea for the spectator expansion of multiple scattering theory used to calculate the effective
interaction employed in elastic $NA$ scattering is an ordering of the scattering process according to the
number of active target nucleons interacting directly with the projectile. In this work we consider the
leading order of the spectator expansion. Thus only two active nucleons are considered. For the leading
order term being derived and calculated {\it ab initio} means that the $NN$ interaction for the active
pair is considered on the same footing as the $NN$ interaction employed to obtain the ground state wave
function and one-body density matrices of the target ground state. Details of the derivation of the
leading order term and how the spin structure of the $NN$ interaction
is consistently taken into account in the reaction process are given in
Refs.~\cite{BurrowsM:2020,Baker:2021izp,Burrows:2020qvu} and shall not be repeated here.   
Since the leading order in the spectator expansion considers two nucleons being active in the
scattering process,
we do not include three-nucleon forces which naturally occur in the chiral expansion starting at
next-to-next-to-leading order (N2LO).

For the construction of the leading order effective interaction, neutron-proton and neutron-neutron 
Wolfenstein amplitudes for a given $NN$ interaction are folded with the nonlocal one-body density
matrices computed within the NCSM framework using the same $NN$ interaction as input. The studies in this
work are based on the semi-local chiral $NN$ interaction by Epelbaum, 
Krebs, and Mei\ss ner~\cite{Epelbaum:2014sza,Epelbaum:2014efa} with a local cutoff $R=1.0$~fm.  
Since the chiral interaction enters on the same footing in the nonlocal $NN$ amplitudes and one-body
density matrices, we can explore the effects of the truncations in the chiral orders of the
interactions in a consistent fashion on the observables of elastic $NA$ scattering.

To quantify the truncation uncertainty arising from each order in the
chiral EFT in the observables we are following two different approaches.
 The first is a pointwise approach, which we apply to bulk quantities such as
 reaction and total cross sections at a specified energy. The second is a correlated approach, which we apply 
to observables that are functions of the scattering angle (momentum transfer) such as the 
differential cross section $d\sigma/d\Omega$, the analyzing power $A_y$, and the spin rotation function $\mathcal{Q}$ at specific energies. In
the following subsections we summarize the most
important features of both approaches. However, we refer the reader to 
Refs.~\cite{Melendez:2017phj, Melendez:2019izc} for detailed descriptions and derivations.

Motivated by the idea of power counting in a chiral EFT, both approaches assume that a quantity $y(x)$ at a given order $k$ can be factorized as
\begin{eqnarray}
y_k(x) = y_{\mathrm{ref}}(x) \sum_{n=0}^k c_n(x) Q^n,
\label{eqn:yk}
\end{eqnarray}
where $y_{\mathrm{ref}}(x)$ is a reference value that includes the dimensions of the quantity $y(x)$ and sets the scale of the problem. The $c_n(x)$ are dimensionless coefficients and $Q$ is the dimensionless expansion parameter for the EFT. We note that in chiral EFT there 
is no term linear in $Q$, i.e.~$c_1\equiv0$. Most chiral EFTs are constructed such that
\begin{eqnarray}
Q = \max \left ( \frac{p}{\Lambda_b}, \frac{M_{\pi}}{\Lambda_b} \right ),
\label{eqn:Q_def}
\end{eqnarray}
where $M_{\pi}$ is the mass of the pion, $\Lambda_b$ is the breakdown scale for the EFT, and $p$ is the relevant momentum. The chiral $NN$ interaction we employ
in this work~\cite{Epelbaum:2014efa} has a breakdown scale of $\Lambda_b=600$~MeV. In previous work
 that studied observables for the $NN$~\cite{Melendez:2017phj} and $Nd$~\cite{Epelbaum:2019zqc} systems, several different choices for the relevant momentum were made. Similar to the previous work in the $Nd$ system, we choose to use the center-of-mass (c.m.) momentum 
of the nucleon-nucleus system $p_{NA}$, written as 
\begin{eqnarray}
p^2_{NA} = \frac{E_{\mathrm{lab}} A^2 m^2 (E_{\mathrm{lab}} + 2m)}{m^2 (A+1)^2 + 2 A m E_{\mathrm{lab}}},
\label{eqn:pnA}
\end{eqnarray}
where $m$ is the nucleon mass, $A$ is the mass number of the target nucleus, and $E_{\mathrm{lab}}$ is the projectile kinetic energy in the laboratory frame. We assess this choice for the expansion parameter in the section
discussing our results by calculating the posteriors for $Q$.

\subsection{Pointwise uncertainty quantification}
The pointwise approach starts from Eq.~(\ref{eqn:yk}) and assumes the quantity of interest $y_k$ is a scalar rather than a functional quantity. This allows to estimate the truncation uncertainty independent of values at nearby $x$ points. This approach is well suited for calculations of reaction and total cross sections, which are calculated  for specific values of the projectile kinetic energy $E_{\mathrm{lab}}$ without regard to the value at nearby energies. Assuming the expansion parameter $Q$ and a reference scale $y_{\mathrm{ref}}$ are known, the pointwise approach uses Eq.~(\ref{eqn:yk}) and the values $y_k$ at each order to extract the coefficients $c_n$. Treating these coefficients as independent draws from the same underlying distribution with variance $\bar{c}^2$,
and we can assign priors based on these beliefs and use Bayes' theorem to show that the posterior distribution for the full prediction is given as
\begin{eqnarray}
\mathrm{pr} (y | \vec{y}_k, Q ) \sim t_{\nu} \left ( y_k, y_{\mathrm{ref}}^2 \frac{Q^{2(k+1)}}{1-Q^2} \tau^2 \right ).
\label{eqn:t_dist}
\end{eqnarray}
This is a student-$t$ distribution with degrees of freedom $\nu$ and scale $\tau$, which gets multiplied by relevant factors in our problem. As many statistical packages have built-in $t$ distributions, confidence intervals corresponding to our truncation uncertainty can be easily calculated using  Eq.~(\ref{eqn:t_dist}).

\subsection{Correlated uncertainty quantification}
The correlated approach also starts from Eq.~(\ref{eqn:yk}), but treats $y_k(x)$ as a functional quantity, and encodes information about nearby $x$ points through a correlation length $\ell$. This approach is well suited for quantities such as the differential cross section, which can be expressed as a function of either the c.m.~angle $\theta_{c.m.}$ or equivalently the momentum transfer $q$. To incorporate information about nearby angles, this approach treats the $c_n$ as independent draws from an underlying Gaussian process $\mathcal{GP}[\mu, \bar{c}^2 r(x,x';\ell)]$. The Gaussian process (GP) is defined with two inputs: a mean function (here we have assumed it is a constant $\mu$) and a covariance function. Following the example of others \cite{Melendez:2019izc}, we have chosen a squared exponential as our covariance function, which factorizes into the variance in the coefficients $\bar{c}^2$ multiplied by a correlation function $r(x,x'; \ell)$, where $\ell$ is the correlation length. 

The GP approach requires training and testing data in order to estimate the correlation length and subsequently produce reliable truncation uncertainty bands. We divide the results into smaller subsets of training versus testing data to accommodate this and employ the model-checking diagnostics introduced in Ref.~\cite{Melendez:2019izc} to assess the quality of our GP fits. The first check is  comparing the true coefficient curves to their corresponding GP emulated curves and assessing if the true curves are properly captured by the emulator. Secondly, we calculate the Mahalanobis distance $D^2_{\mathrm{MD}}$, which is a multivariate analog to the idea of calculating the sum of the squared residuals to measure loss. The quantity $D_{\mathrm{MD}}^2$ takes into account the correlations our GP builds in.
A  large value of $D^2_{\mathrm{MD}}$ implies the emulator is not reproducing the validation data. Lastly, we also calculate a pivoted Cholesky decomposition $D_{\mathrm{PC}}$, which can identify the data that is contributing to a failing $D^2_{\mathrm{MD}}$. Patterns in the $D_{\mathrm{PC}}$ values when plotted versus index can indicate variances or correlation lengths that have been incorrectly estimated \cite{Melendez:2019izc}. We also assess the choice of the expansion parameter $Q$ by calculating the marginal posterior for $Q$ and comparing its maximum \textit{a posteriori} (MAP) value with the value attained by using Eqs.~(\ref{eqn:Q_def}) and (\ref{eqn:pnA}). We
refer the reader to Eqs.~(A49) and (A53) in Ref.~\cite{Melendez:2019izc} for more details.

%% file: results.tex

\subsection{Reaction and total cross sections}
Reaction and total cross sections are represented by a single number at each projectile energy, and
are therefore best suited to apply the pointwise, uncorrelated approach to uncertainty quantification.
For proton scattering from $^{16}$O we consider the reaction cross section and for neutron scattering
from $^{12}$C the total cross section. Since the construction of the effective $NA$ interaction requires
the folding of a one-body density matrix obtained from NCSM calculations with the $NN$ interaction
calculated at the same order of the chiral expansion, we not only have chiral truncation errors but
also numerical errors coming from the corresponding NCSM structure calculation.  

As examples we choose the reaction cross section of  $^{16}$O$(p,p)^{16}$O at 100~MeV projectile
kinetic energy and the total cross section of $^{12}$C$(n,n)^{12}$C at 95~MeV, and study the effect of
the truncation errors in the chiral EFT as well as the numerical uncertainty of the NCSM calculation. 
For the reaction cross section (Fig.~\ref{fig1}(a)), we see the estimated $1\sigma$ truncation uncertainty bands are larger than the numerical uncertainty associated with $\hbar\Omega$ at each $N_{\mathrm{max}}$ value. As $N_{\mathrm{max}}$ increases, the range of possible values resulting from changes in $\hbar\Omega$ decreases slightly, but at $N_{\mathrm{max}}=10$ the truncation uncertainty dominates the overall uncertainty in this quantity.

For the total cross section, the inset of Fig.~\ref{fig1}(b) shows a similar behavior of truncation uncertainty versus numerical uncertainty as was shown for the reaction cross section in Fig.~\ref{fig1}(a). The total cross section for neutron scattering on $^{12}$C has been studied in the experimental literature and we have included those values and error bars in Fig.~\ref{fig1}(b). At N2LO, those experimental values fall within the $1\sigma$ truncation uncertainty. To estimate higher order effects on $\sigma_{\mathrm{tot}}$, we have also performed inconsistent calculations in which the $NN$
amplitudes at N3LO or N4LO are used to calculate the effective interaction, but they are combined with the N2LO one-body density matrix in each case. This is indicated in the figure by empty circles at N3LO and N4LO. Including these results in the uncertainty quantification does provide smaller $1\sigma$ uncertainty bands, though we note the experimental values will still fall within the $2\sigma$ bands.

\subsection{Differential cross section}
Next, we consider functional quantities, i.e.~observables that depend on the momentum transfer (or
equivalently on the scattering angle), and concentrate first on differential cross sections.
To estimate the truncation uncertainty in functional quantities, we use the correlated approach which relies on Gaussian processes. The differential cross section divided by the Rutherford cross
section is calculated for proton scattering from $^{16}$O at various projectile kinetic energies and compared to experimental data
as is shown in Fig.~\ref{fig2}. These results indicate a strong dependence on the expansion parameter $Q$ at higher energies, as
is expected by the tabulated values of $Q$ shown in Table \ref{tab1}. Specifically, the truncation uncertainty bands at N2LO for 180 MeV
projectile kinetic energy are sufficiently large that the predictive power at that energy is virtually nonexistent. Nonetheless, the increasing agreement with data in the first peak and first minimum as higher orders are included gives
the correct trend. Minima in the differential cross section correlate with the size of the target nucleus. It
is well known \cite{Binder:2018pgl} that the nuclear binding energy calculated with the LO of the
chiral $NN$ interaction is way too large and correspondingly the radius much too small. Only when going
to NLO and N2LO the binding energy as well as the radius move into the vicinity of their experimental
values. This insight from structure calculations is corroborated by the calculations in
Fig.~\ref{fig2}, where with increasing chiral order the calculated first diffraction minimum moves
towards smaller angles (momentum transfers) indicating a larger nuclear size.

We further apply the same approach to the differential cross section for neutron scattering from $^{12}$C. 
The calculations are shown in Fig.~\ref{fig3}. Here the angular range is chosen to only cover the range
for which data are available. Considering both the experimental error bars and the truncation error bars, we see good agreement with the available data. Similar to the proton scattering case, we again see the effect of the large expansion parameter at higher energies, with the truncation error bars remaining large at N2LO.

To assess our choice of the procedure to estimate the expansion parameter, we calculated posteriors for $Q$ given the $^{16}$O$(p,p)^{16}$O and $^{12}$C$(n,n)^{12}$C differential cross sections at each energy (Fig.~\ref{fig4}). From each of these posteriors, we can extract the maximum \textit{a posteriori} (MAP) value corresponding to the single value best guess for that quantity. Comparing this MAP value with the prescription for $Q$ we have implemented, we can see they are in generally good agreement though there is some freedom to choose smaller or larger $Q$ values. It is worthwhile to note the differential cross sections shown here are truncated such that values corresponding to a momentum transfer $q$ larger than $p_{NA}$ are excluded to alleviate any concerns about it shifting $Q$ to even larger values. In
addition, for proton scattering, we excluded the smallest angles in
the differential cross section because they are dominated by Rutherford scattering.
This was done in order to assess the truncation errors arising solely from the nuclear interaction and achieve good fits
for the Gaussian process.

Examples of the Gaussian process diagnostics used to assess our fits for proton and neutron scattering are shown in Figs.~\ref{fig5} and \ref{fig6}, respectively. As previously mentioned, for proton scattering, the Gaussian process ignores small angles when fitting because the effect of Rutherford scattering rapidly alters the correlation length. This behavior yields bad diagnostics, particularly for the Mahalanobis distance $D_{\mathrm{MD}}^2$, which increases to unrealistically large values. Choosing to not train or test the GP in that region alleviates those concerns -- this is indicated in Fig.~\ref{fig5} by a lack of tick marks at small angles. In contrast, the GP for neutron scattering can be trained and tested on small angles while still yielding realistic values for the $D^2_{\mathrm{MD}}$. Unlike the GP applications in $NN$ scattering where $\mu=0$, we find a small, positive, constant mean $\mu$ yields slightly better fits, especially for the pivoted Cholesky decomposition $D_{\mathrm{PC}}$. In both Figs.~\ref{fig5} and \ref{fig6}, we used $\mu=0.5$, which make the $D_{\mathrm{PC}}$ values more equally distributed among postive and negative values. If $\mu=0$ is used instead, the $D_{\mathrm{PC}}$ values are noticeably more positive than negative. While this mean value was determined empirically, future work could learn an appropriate mean from the data, much like what was done for the expansion parameter.

\subsection{Analyzing power}
Spin observables usually give more detailed insights into the effective $NA$ interaction, since they
are ratios between cross sections and absolute magnitudes are divided out. The scattering of a spin-$\frac{1}{2}$
proton from a spin-$0$ nucleus allows for two independent spin observables, the analyzing power $A_y$ for
transverse polarized protons, and the spin rotation function $\mathcal Q$ for longitudinal polarized
protons. 

In this subsection we want to concentrate on $A_y$ for
proton scattering from  $^{16}$O at a selection of laboratory kinetic energies.
In Fig.~\ref{fig7} calculations based on LO, NLO, and N2LO in the chiral $NN$ interaction are
shown for 65~MeV, 100~MeV, and 135~MeV and compared to available experimental data.  Chiral $NN$
interactions only acquire spin-orbit contributions in NLO, which is clearly seen in the middle column
in Fig.~\ref{fig7}, where the calculations start to follow the structure of the data.
We see good agreement between the theoretical results and experimental data at 100~MeV, particularly in forward directions, but observe noticeable differences between theory at N2LO and experimental data at 135~MeV. 
Connecting this with observations for the differential cross section at higher energies suggests that the chiral interaction
we employ for the current study may be best suited to describe experimental scattering results at 
projectile energies around 100~MeV or lower. To test if this observation is independent of the choice of
nucleus, we
substitute proton scattering from $^{12}$C for the equivalent neutron scattering calculations.
In Fig.~\ref{fig8} the analyzing power for proton scattering on $^{12}$C at 65  and 122~MeV is
shown. Similar to $^{16}$O, these results show good agreement at forward angles, with differences between theory and experiment developing at larger angles/higher momentum transfers.

The associated GP diagnostics for the analyzing power in proton scattering from $^{16}$O are shown in Fig.~\ref{fig9}. The gray dashed line in the coefficient plot, which goes to zero at $0\degree$, indicates we have used the symmetry-constrained GP procedure from Ref.~\cite{Melendez:2019izc}. Since the value of the $A_y$ must be zero at that point, the truncation error bars should also go to zero. All of our plots in Fig.~\ref{fig8} start at $2\degree$, and because the value of the error bars grows rapidly, this effect cannot be seen there. It should be noted that the Mahalanobis distance $D^2_{\mathrm{MD}}$ for LO is essentially zero and lies outside the $95\%$ CI represented by the the whiskers on the box plot, indicating that the LO coefficient curve may not be well-captured by the GP. Since the LO result is essentially zero at all angles (momentum transfers), this matches our expectation and others observations \cite{Melendez:2017phj} that the leading order result may not be informative to our analysis, particularly for spin observables. Lastly, again, the analyzing power required a small nonzero mean ($\mu=0.3$) to equally distribute the $D_{\mathrm{PC}}$ values.

\subsection{Spin rotation function}
The spin rotation function ${\mathcal Q}$ is the second, independent spin observable in scattering of
protons from spin-0 nuclei.  Experimental information for this observable is considerably scarcer, 
since its determination requires analyzing the polarization of the scattered 
particles~\cite{wolfenstein-1954}.  
Fortunately experimental information is available at 65~MeV for proton scattering
from $^{16}$O and $^{12}$C. In Fig.~\ref{fig10} we present our calculations together with the
experimental data.

In both cases, the N2LO result with its associated truncation error bars captures most of the data, particularly in more forward directions.
We want to point out, that already the NLO calculations follow the general shape of the data. As in the
case of the analyzing power, the LO calculation for which the chiral interaction lacks spin-operators
shows a zero spin rotation function in forward direction. 

The associated GP diagnostics (Fig.~\ref{fig11}) illustrate that the coefficient curves for the spin rotation function are more difficult to model than the previous examples, but decent fits can still be obtained. Notably, here, the $D^2_{\mathrm{MD}}$ values for each order are more spread out than previous cases, though only the LO value falls outside of the $95\%$ confidence interval indicated by the whiskers on the box plot. This is similar to the behavior seen for the analyzing power and again may be a reflection of the leading order result not being informative for spin observables. Unlike both the differential cross section and the analyzing power, the pivoted Cholesky decomposition for the spin rotation function did not require the use of a nonzero mean to evenly distribute its points across the $x$-axis. This may be a reflection of the behavior of the result at each order for this particular interaction -- that is, in the case of the differential cross sections and the analyzing powers we have calculated here, the result at each order is almost always larger (more positive) than the result at the previous order. In constrast, the values for the spin rotation function are essentially zero at LO, become negative at NLO, and then become less negative at N2LO. This oscillating behavior between orders may have alleviated the need for a nonzero mean when fitting.

\subsection{Neutron-nucleus scattering in forward direction}
The scattering observables and their analysis in terms of the truncation uncertainties coming 
from the underlying chiral
EFT indicate that when looking at small momentum transfer, there seems to be a similar behavior in the
convergence pattern as was observed in studies of the $NN$
observables~\cite{Epelbaum:2014sza,Epelbaum:2014efa,Furnstahl:2015rha}.
This is a nontrivial observation, since \textit{a priori} one can not expect that e.g.~multiple scattering
effects resulting from first solving the integral equation to obtain the Watson
potential~\cite{Chinn:1993zza} and then second from solving the Lippmann-Schwinger integral equation
do not influence the analysis. 
Our calculations are based on an {\it ab initio} effective interaction calculated in the leading of the
spectator expansion of a multiple scattering theory~\cite{Burrows:2020qvu}, in which a one-body nuclear
density matrix is folded with $NN$ amplitudes calculated from the same $NN$ force. 
In the elastic scattering
of a proton (neutron) from a spin-0 nucleus the effective interaction contains only two contributions, a
spin-independent central potential and a spin-orbit potential. As shown in detail in 
Refs.~\cite{Burrows:2020qvu,BurrowsM:2020}, the central potential is built by folding 
the $NN$ Wolfenstein~\cite{wolfenstein-ashkin} amplitudes $A$ and $C$ with the nuclear density matrix, 
while the spin-orbit potential contains contributions from the $NN$ Wolfenstein amplitudes $C$ and $M$. 
Furthermore, one should note that the
amplitude $C$ is zero for zero-momentum transfer, and very small for momentum transfers
$q \leq 1$~fm$^{-1}$. This means that the small momentum region is dominated by the
spin-independent components of the neutron-proton ($np$) and proton-proton ($pp$) amplitudes $A$ and thus
studying the elastic $NA$ differential cross section in this region opens a unique window on specific pieces of the $NN$
interaction.

In order to investigate if there is a direct correlation between the Wolfenstein amplitude $A$
and the differential cross sections for neutron scattering at small momentum transfers, we should
consider the order-by-order contributions to the square of $A$, summed over $np$ and $pp$
contributions. In Fig.~\ref{fig12}(a) this quantity is plotted as a function of the momentum transfer in
the $NN$ system at the energy entering the neutron-$^{12}$C scattering (95~MeV), where the result at each
order of the chiral expansion is shown. As expected from
Refs.~\cite{Epelbaum:2014sza,Epelbaum:2014efa}, the result shows excellent
convergence with increasing chiral order. In Fig.~\ref{fig12}(b) the differential cross section for
neutron scattering from $^{12}$C is shown for the same momentum transfer based on the same chiral
orders.  We note that consistent {\it ab initio} calculations are only carried out up to N2LO. For the
N3LO and N4LO calculations we used the one-body densities obtained in N2LO and folded them with
the corresponding $NN$ amplitudes in the higher chiral orders. 
We expect this inconsistency to have little effect on the differential cross section in forward
direction in our investigation, 
since the main features of the ground state one-body density matrix are already established
at the N2LO level. We observe a very similar behavior in the convergence with respect to the chiral
order as seen in Fig.~\ref{fig12}(a). 
The qualitative similarities between the two quantities are striking, with the values of both increasing as higher order contributes are included. From N2LO
on, the changes induced by N3LO and N4LO contributions are small in comparison. 

To explore the connection between these two quantities further, we have plotted them as functions of each other in Fig.~\ref{fig13}(a) for momentum transfers
$q \le 0.55$~fm$^{-1}$.
This results in a linear correlation, albeit with slightly differing slopes and $y$-intercepts. 
Normalizing for these differences using the same technique as discussed in Ref.~\cite{LauneySDD_CPC14} and implemented in Refs.~\cite{Burrows:2018ggt, Baker:2021izp}, in Fig.~\ref{fig13}(b) we see these two quantities are strongly correlated with a correlation coefficient of $0.99$. This implies that the forward direction in $NA$ scattering provides a direct connection to the underlying $NN$ interaction. We
also found that a simliarly good correlation between the squares of the Wolfenstein amplitudes $A$ and
the neutron differential cross section exists for energies as low as 65~MeV and as high as 185~MeV.
This observation may be useful when attempting to link properties of a $NN$ interaction to its effects
in elastic $NA$ observables, even if only in a small range of momentum transfers.

%% file: conclusions.tex

We have successfully implemented two procedures to quantify the theoretical uncertainties associated with the underlying chiral EFT used to describe \textit{ab initio} nucleon-nucleus elastic scattering. While both procedures have been applied in the $NN$ and $Nd$ systems, we extend their application to the $NA$ system to study proton scattering from $^{16}$O and neutron scattering from $^{12}$C at various energies. For both reactions, we have shown that the uncertainty associated with the chiral expansion is larger than the numerical uncertainty associated with the many-body method used to describe the target nucleus. We also have shown that our prescription for estimating the expansion parameter using the nucleon-nucleus center-of-mass momentum is supported by an analysis of the posterior distributions for $Q$. This places a limit on the energy range where we can apply these tools because as the projectile energies approach 200 MeV, this yields a chiral EFT expansion parameter of approximately one. 
At energies of about 100~MeV projectile kinetic energy and lower these tools work
very well, showing that with increasing chiral order the truncation uncertainty decreases. We also find that
the chiral $NN$ interaction we base our studies on~\cite{Epelbaum:2014sza,Epelbaum:2014efa} agrees very
well with experiment at 100~MeV and lower. At energies higher than 100~MeV, the expansion parameter increases and as a result
the chiral truncation uncertainties become very large. We also see a deterioration of the
agreement of our central values with experiment for those energies. This is somewhat in contrast to previous studies and
experiences using different interactions~\cite{Burrows:2020qvu,Weppner:1997wx,Elster:1996xh} which described
$NN$ observables about equally well at energies up to 200 MeV. The systematic study of chiral truncation
uncertainties, carried out here for the first time, seems to indicate that {\it ab initio} effective
$NA$ interactions derived from certain chiral EFTs allows for a good description of experiment at
energies lower than previously assumed, provided we focus on regions of momentum transfer where the
analysis of the EFT truncation uncertainty is valid. 

While examining the differential cross sections in forward directions and low momentum transfers, both proton-nucleus and neutron-nucleus reactions yielded useful insights. For proton-nucleus scattering, we saw that the influence of Rutherford scattering can negatively affect the tools used for uncertainty quantification as the effects coming from the nuclear interaction are overwhelmed by contributions from the Coulomb interaction. For neutron-nucleus scattering, we have shown that the forward direction can provide a unique window into the underlying $NN$ interaction. Specifically, we identified a strong correlation between the differential cross section in $^{12}$C$(n,n)^{12}$C and the 
 $NN$ Wolfenstein amplitude $A$ representing the spin-independent part of the $NN$ interaction at low momentum transfers. Thus, provided there is experimental data for neutron-nucleus scattering at low momentum transfers, this could indicate a new region to examine when 
assessing a $NN$ interaction.

%% file: order_by_order_paper.bbl
\begin{thebibliography}{44}%
\makeatletter
\providecommand \@ifxundefined [1]{%
 \@ifx{#1\undefined}
}%
\providecommand \@ifnum [1]{%
 \ifnum #1\expandafter \@firstoftwo
 \else \expandafter \@secondoftwo
 \fi
}%
\providecommand \@ifx [1]{%
 \ifx #1\expandafter \@firstoftwo
 \else \expandafter \@secondoftwo
 \fi
}%
\providecommand \natexlab [1]{#1}%
\providecommand \enquote  [1]{``#1''}%
\providecommand \bibnamefont  [1]{#1}%
\providecommand \bibfnamefont [1]{#1}%
\providecommand \citenamefont [1]{#1}%
\providecommand \href@noop [0]{\@secondoftwo}%
\providecommand \href [0]{\begingroup \@sanitize@url \@href}%
\providecommand \@href[1]{\@@startlink{#1}\@@href}%
\providecommand \@@href[1]{\endgroup#1\@@endlink}%
\providecommand \@sanitize@url [0]{\catcode `\\12\catcode `\$12\catcode
  `\&12\catcode `\#12\catcode `\^12\catcode `\_12\catcode `\%12\relax}%
\providecommand \@@startlink[1]{}%
\providecommand \@@endlink[0]{}%
\providecommand \url  [0]{\begingroup\@sanitize@url \@url }%
\providecommand \@url [1]{\endgroup\@href {#1}{\urlprefix }}%
\providecommand \urlprefix  [0]{URL }%
\providecommand \Eprint [0]{\href }%
\providecommand \doibase [0]{http://dx.doi.org/}%
\providecommand \selectlanguage [0]{\@gobble}%
\providecommand \bibinfo  [0]{\@secondoftwo}%
\providecommand \bibfield  [0]{\@secondoftwo}%
\providecommand \translation [1]{[#1]}%
\providecommand \BibitemOpen [0]{}%
\providecommand \bibitemStop [0]{}%
\providecommand \bibitemNoStop [0]{.\EOS\space}%
\providecommand \EOS [0]{\spacefactor3000\relax}%
\providecommand \BibitemShut  [1]{\csname bibitem#1\endcsname}%
\let\auto@bib@innerbib\@empty
\bibitem [{\citenamefont {Epelbaum}\ \emph
  {et~al.}(2015{\natexlab{a}})\citenamefont {Epelbaum}, \citenamefont {Krebs},\
  and\ \citenamefont {Mei\ss{}ner}}]{Epelbaum:2014sza}%
  \BibitemOpen
  \bibfield  {author} {\bibinfo {author} {\bibfnamefont {E.}~\bibnamefont
  {Epelbaum}}, \bibinfo {author} {\bibfnamefont {H.}~\bibnamefont {Krebs}}, \
  and\ \bibinfo {author} {\bibfnamefont {U.~G.}\ \bibnamefont {Mei\ss{}ner}},\
  }\href {\doibase 10.1103/PhysRevLett.115.122301} {\bibfield  {journal}
  {\bibinfo  {journal} {Phys. Rev. Lett.}\ }\textbf {\bibinfo {volume} {115}},\
  \bibinfo {pages} {122301} (\bibinfo {year} {2015}{\natexlab{a}})},\ \Eprint
  {http://arxiv.org/abs/1412.4623} {arXiv:1412.4623 [nucl-th]} \BibitemShut
  {NoStop}%
\bibitem [{\citenamefont {Epelbaum}\ \emph
  {et~al.}(2015{\natexlab{b}})\citenamefont {Epelbaum}, \citenamefont {Krebs},\
  and\ \citenamefont {Mei\ss{}ner}}]{Epelbaum:2014efa}%
  \BibitemOpen
  \bibfield  {author} {\bibinfo {author} {\bibfnamefont {E.}~\bibnamefont
  {Epelbaum}}, \bibinfo {author} {\bibfnamefont {H.}~\bibnamefont {Krebs}}, \
  and\ \bibinfo {author} {\bibfnamefont {U.~G.}\ \bibnamefont {Mei\ss{}ner}},\
  }\href {\doibase 10.1140/epja/i2015-15053-8} {\bibfield  {journal} {\bibinfo
  {journal} {Eur. Phys. J. A}\ }\textbf {\bibinfo {volume} {51}},\ \bibinfo
  {pages} {53} (\bibinfo {year} {2015}{\natexlab{b}})},\ \Eprint
  {http://arxiv.org/abs/1412.0142} {arXiv:1412.0142 [nucl-th]} \BibitemShut
  {NoStop}%
\bibitem [{\citenamefont {Reinert}\ \emph {et~al.}(2018)\citenamefont
  {Reinert}, \citenamefont {Krebs},\ and\ \citenamefont
  {Epelbaum}}]{Reinert:2017usi}%
  \BibitemOpen
  \bibfield  {author} {\bibinfo {author} {\bibfnamefont {P.}~\bibnamefont
  {Reinert}}, \bibinfo {author} {\bibfnamefont {H.}~\bibnamefont {Krebs}}, \
  and\ \bibinfo {author} {\bibfnamefont {E.}~\bibnamefont {Epelbaum}},\ }\href
  {\doibase 10.1140/epja/i2018-12516-4} {\bibfield  {journal} {\bibinfo
  {journal} {Eur. Phys. J. A}\ }\textbf {\bibinfo {volume} {54}},\ \bibinfo
  {pages} {86} (\bibinfo {year} {2018})},\ \Eprint
  {http://arxiv.org/abs/1711.08821} {arXiv:1711.08821 [nucl-th]} \BibitemShut
  {NoStop}%
\bibitem [{\citenamefont {Epelbaum}\ \emph
  {et~al.}(2020{\natexlab{a}})\citenamefont {Epelbaum}, \citenamefont {Krebs},\
  and\ \citenamefont {Reinert}}]{Epelbaum:2019kcf}%
  \BibitemOpen
  \bibfield  {author} {\bibinfo {author} {\bibfnamefont {E.}~\bibnamefont
  {Epelbaum}}, \bibinfo {author} {\bibfnamefont {H.}~\bibnamefont {Krebs}}, \
  and\ \bibinfo {author} {\bibfnamefont {P.}~\bibnamefont {Reinert}},\ }\href
  {\doibase 10.3389/fphy.2020.00098} {\bibfield  {journal} {\bibinfo  {journal}
  {Front. in Phys.}\ }\textbf {\bibinfo {volume} {8}},\ \bibinfo {pages} {98}
  (\bibinfo {year} {2020}{\natexlab{a}})},\ \Eprint
  {http://arxiv.org/abs/1911.11875} {arXiv:1911.11875 [nucl-th]} \BibitemShut
  {NoStop}%
\bibitem [{\citenamefont {Machleidt}\ and\ \citenamefont
  {Entem}(2011)}]{Machleidt:2011zz}%
  \BibitemOpen
  \bibfield  {author} {\bibinfo {author} {\bibfnamefont {R.}~\bibnamefont
  {Machleidt}}\ and\ \bibinfo {author} {\bibfnamefont {D.~R.}\ \bibnamefont
  {Entem}},\ }\href {\doibase 10.1016/j.physrep.2011.02.001} {\bibfield
  {journal} {\bibinfo  {journal} {Phys. Rept.}\ }\textbf {\bibinfo {volume}
  {503}},\ \bibinfo {pages} {1} (\bibinfo {year} {2011})},\ \Eprint
  {http://arxiv.org/abs/1105.2919} {arXiv:1105.2919 [nucl-th]} \BibitemShut
  {NoStop}%
\bibitem [{\citenamefont {Entem}\ \emph {et~al.}(2017)\citenamefont {Entem},
  \citenamefont {Machleidt},\ and\ \citenamefont {Nosyk}}]{Entem:2017gor}%
  \BibitemOpen
  \bibfield  {author} {\bibinfo {author} {\bibfnamefont {D.~R.}\ \bibnamefont
  {Entem}}, \bibinfo {author} {\bibfnamefont {R.}~\bibnamefont {Machleidt}}, \
  and\ \bibinfo {author} {\bibfnamefont {Y.}~\bibnamefont {Nosyk}},\ }\href
  {\doibase 10.1103/PhysRevC.96.024004} {\bibfield  {journal} {\bibinfo
  {journal} {Phys. Rev. C}\ }\textbf {\bibinfo {volume} {96}},\ \bibinfo
  {pages} {024004} (\bibinfo {year} {2017})},\ \Eprint
  {http://arxiv.org/abs/1703.05454} {arXiv:1703.05454 [nucl-th]} \BibitemShut
  {NoStop}%
\bibitem [{\citenamefont {Langr}\ \emph
  {et~al.}(2019{\natexlab{a}})\citenamefont {Langr}, \citenamefont {Dytrych},
  \citenamefont {Draayer}, \citenamefont {Launey},\ and\ \citenamefont
  {Tvrdik}}]{LangrDDLT19}%
  \BibitemOpen
  \bibfield  {author} {\bibinfo {author} {\bibfnamefont {D.}~\bibnamefont
  {Langr}}, \bibinfo {author} {\bibfnamefont {T.}~\bibnamefont {Dytrych}},
  \bibinfo {author} {\bibfnamefont {J.~P.}\ \bibnamefont {Draayer}}, \bibinfo
  {author} {\bibfnamefont {K.~D.}\ \bibnamefont {Launey}}, \ and\ \bibinfo
  {author} {\bibfnamefont {P.}~\bibnamefont {Tvrdik}},\ }\href {\doibase
  10.1016/j.cpc.2019.05.018} {\bibfield  {journal} {\bibinfo  {journal} {Comp.
  Phys. Comm.}\ }\textbf {\bibinfo {volume} {244}},\ \bibinfo {pages} {442}
  (\bibinfo {year} {2019}{\natexlab{a}})}\BibitemShut {NoStop}%
\bibitem [{\citenamefont {Langr}\ \emph
  {et~al.}(2019{\natexlab{b}})\citenamefont {Langr}, \citenamefont {Dytrych},
  \citenamefont {Launey},\ and\ \citenamefont {Draayer}}]{LangrDDT18}%
  \BibitemOpen
  \bibfield  {author} {\bibinfo {author} {\bibfnamefont {D.}~\bibnamefont
  {Langr}}, \bibinfo {author} {\bibfnamefont {T.}~\bibnamefont {Dytrych}},
  \bibinfo {author} {\bibfnamefont {K.~D.}\ \bibnamefont {Launey}}, \ and\
  \bibinfo {author} {\bibfnamefont {J.~P.}\ \bibnamefont {Draayer}},\ }\href
  {\doibase 10.1177/1094342019838314} {\bibfield  {journal} {\bibinfo
  {journal} {The Int. J. of High Performance Computing Applications}\ }\textbf
  {\bibinfo {volume} {33}},\ \bibinfo {pages} {522} (\bibinfo {year}
  {2019}{\natexlab{b}})}\BibitemShut {NoStop}%
\bibitem [{\citenamefont {Shao}\ \emph {et~al.}(2018)\citenamefont {Shao},
  \citenamefont {Aktulga}, \citenamefont {Yang}, \citenamefont {Ng},
  \citenamefont {Maris},\ and\ \citenamefont {Vary}}]{SHAO20181}%
  \BibitemOpen
  \bibfield  {author} {\bibinfo {author} {\bibfnamefont {M.}~\bibnamefont
  {Shao}}, \bibinfo {author} {\bibfnamefont {H.~M.}\ \bibnamefont {Aktulga}},
  \bibinfo {author} {\bibfnamefont {C.}~\bibnamefont {Yang}}, \bibinfo {author}
  {\bibfnamefont {E.~G.}\ \bibnamefont {Ng}}, \bibinfo {author} {\bibfnamefont
  {P.}~\bibnamefont {Maris}}, \ and\ \bibinfo {author} {\bibfnamefont {J.~P.}\
  \bibnamefont {Vary}},\ }\href {\doibase
  https://doi.org/10.1016/j.cpc.2017.09.004} {\bibfield  {journal} {\bibinfo
  {journal} {Comp. Phys. Comm.}\ }\textbf {\bibinfo {volume} {222}},\ \bibinfo
  {pages} {1 } (\bibinfo {year} {2018})},\ \Eprint
  {http://arxiv.org/abs/1609.01689} {arXiv:1609.01689 [nucl-th]} \BibitemShut
  {NoStop}%
\bibitem [{\citenamefont {Aktulga}\ \emph {et~al.}(2014)\citenamefont
  {Aktulga}, \citenamefont {Yang}, \citenamefont {Ng}, \citenamefont {Maris},\
  and\ \citenamefont {Vary}}]{CPE:CPE3129}%
  \BibitemOpen
  \bibfield  {author} {\bibinfo {author} {\bibfnamefont {H.~M.}\ \bibnamefont
  {Aktulga}}, \bibinfo {author} {\bibfnamefont {C.}~\bibnamefont {Yang}},
  \bibinfo {author} {\bibfnamefont {E.~G.}\ \bibnamefont {Ng}}, \bibinfo
  {author} {\bibfnamefont {P.}~\bibnamefont {Maris}}, \ and\ \bibinfo {author}
  {\bibfnamefont {J.~P.}\ \bibnamefont {Vary}},\ }\href {\doibase
  10.1002/cpe.3129} {\bibfield  {journal} {\bibinfo  {journal} {Concurrency and
  Computation: Practice and Experience}\ }\textbf {\bibinfo {volume} {26}},\
  \bibinfo {pages} {2631} (\bibinfo {year} {2014})}\BibitemShut {NoStop}%
\bibitem [{\citenamefont {Jung}\ \emph {et~al.}(2013)\citenamefont {Jung},
  \citenamefont {Wilson}, \citenamefont {Choi}, \citenamefont {Shalf},
  \citenamefont {Aktulga}, \citenamefont {Yang}, \citenamefont {Saule},
  \citenamefont {Catalyurek},\ and\ \citenamefont {Kandemir}}]{Jung:2013:EFO}%
  \BibitemOpen
  \bibfield  {author} {\bibinfo {author} {\bibfnamefont {M.}~\bibnamefont
  {Jung}}, \bibinfo {author} {\bibfnamefont {E.~H.}\ \bibnamefont {Wilson},
  \bibfnamefont {III}}, \bibinfo {author} {\bibfnamefont {W.}~\bibnamefont
  {Choi}}, \bibinfo {author} {\bibfnamefont {J.}~\bibnamefont {Shalf}},
  \bibinfo {author} {\bibfnamefont {H.~M.}\ \bibnamefont {Aktulga}}, \bibinfo
  {author} {\bibfnamefont {C.}~\bibnamefont {Yang}}, \bibinfo {author}
  {\bibfnamefont {E.}~\bibnamefont {Saule}}, \bibinfo {author} {\bibfnamefont
  {U.~V.}\ \bibnamefont {Catalyurek}}, \ and\ \bibinfo {author} {\bibfnamefont
  {M.}~\bibnamefont {Kandemir}},\ }in\ \href {\doibase 10.1145/2503210.2503261}
  {\emph {\bibinfo {booktitle} {Proceedings of the International Conference on
  High Performance Computing, Networking, Storage and Analysis}}},\ \bibinfo
  {series and number} {SC '13}\ (\bibinfo  {publisher} {ACM},\ \bibinfo
  {address} {New York, NY, USA},\ \bibinfo {year} {2013})\ pp.\ \bibinfo
  {pages} {75:1--75:11}\BibitemShut {NoStop}%
\bibitem [{\citenamefont {Navratil}\ \emph {et~al.}(2000)\citenamefont
  {Navratil}, \citenamefont {Vary},\ and\ \citenamefont
  {Barrett}}]{Navratil:2000ww}%
  \BibitemOpen
  \bibfield  {author} {\bibinfo {author} {\bibfnamefont {P.}~\bibnamefont
  {Navratil}}, \bibinfo {author} {\bibfnamefont {J.~P.}\ \bibnamefont {Vary}},
  \ and\ \bibinfo {author} {\bibfnamefont {B.~R.}\ \bibnamefont {Barrett}},\
  }\href {\doibase 10.1103/PhysRevLett.84.5728} {\bibfield  {journal} {\bibinfo
   {journal} {Phys. Rev. Lett.}\ }\textbf {\bibinfo {volume} {84}},\ \bibinfo
  {pages} {5728} (\bibinfo {year} {2000})},\ \Eprint
  {http://arxiv.org/abs/nucl-th/0004058} {arXiv:nucl-th/0004058 [nucl-th]}
  \BibitemShut {NoStop}%
\bibitem [{\citenamefont {Roth}\ and\ \citenamefont
  {Navratil}(2007)}]{Roth:2007sv}%
  \BibitemOpen
  \bibfield  {author} {\bibinfo {author} {\bibfnamefont {R.}~\bibnamefont
  {Roth}}\ and\ \bibinfo {author} {\bibfnamefont {P.}~\bibnamefont
  {Navratil}},\ }\href {\doibase 10.1103/PhysRevLett.99.092501} {\bibfield
  {journal} {\bibinfo  {journal} {Phys. Rev. Lett.}\ }\textbf {\bibinfo
  {volume} {99}},\ \bibinfo {pages} {092501} (\bibinfo {year} {2007})},\
  \Eprint {http://arxiv.org/abs/0705.4069} {arXiv:0705.4069 [nucl-th]}
  \BibitemShut {NoStop}%
\bibitem [{\citenamefont {Barrett}\ \emph {et~al.}(2013)\citenamefont
  {Barrett}, \citenamefont {Navr\'{a}til},\ and\ \citenamefont
  {Vary}}]{BarrettNV13}%
  \BibitemOpen
  \bibfield  {author} {\bibinfo {author} {\bibfnamefont {B.}~\bibnamefont
  {Barrett}}, \bibinfo {author} {\bibfnamefont {P.}~\bibnamefont
  {Navr\'{a}til}}, \ and\ \bibinfo {author} {\bibfnamefont {J.}~\bibnamefont
  {Vary}},\ }\href@noop {} {\bibfield  {journal} {\bibinfo  {journal} {Prog.
  Part. Nucl. Phys.}\ }\textbf {\bibinfo {volume} {69}},\ \bibinfo {pages}
  {131} (\bibinfo {year} {2013})}\BibitemShut {NoStop}%
\bibitem [{\citenamefont {Binder}\ \emph {et~al.}(2018)\citenamefont {Binder},
  \citenamefont {Calci}, \citenamefont {Epelbaum}, \citenamefont {Furnstahl},
  \citenamefont {Golak}, \citenamefont {Hebeler}, \citenamefont {H\"{u}ther},
  \citenamefont {Kamada}, \citenamefont {Krebs}, \citenamefont {Maris},
  \citenamefont {Mei\ss{}ner}, \citenamefont {Nogga}, \citenamefont {Roth},
  \citenamefont {Skibi\ifmmode~\acute{n}\else \'{n}\fi{}ski}, \citenamefont
  {Topolnicki}, \citenamefont {Vary}, \citenamefont {Vobig},\ and\
  \citenamefont {Wita\l{}a}}]{Binder:2018pgl}%
  \BibitemOpen
  \bibfield  {author} {\bibinfo {author} {\bibfnamefont {S.}~\bibnamefont
  {Binder}}, \bibinfo {author} {\bibfnamefont {A.}~\bibnamefont {Calci}},
  \bibinfo {author} {\bibfnamefont {E.}~\bibnamefont {Epelbaum}}, \bibinfo
  {author} {\bibfnamefont {R.~J.}\ \bibnamefont {Furnstahl}}, \bibinfo {author}
  {\bibfnamefont {J.}~\bibnamefont {Golak}}, \bibinfo {author} {\bibfnamefont
  {K.}~\bibnamefont {Hebeler}}, \bibinfo {author} {\bibfnamefont
  {T.}~\bibnamefont {H\"{u}ther}}, \bibinfo {author} {\bibfnamefont
  {H.}~\bibnamefont {Kamada}}, \bibinfo {author} {\bibfnamefont
  {H.}~\bibnamefont {Krebs}}, \bibinfo {author} {\bibfnamefont
  {P.}~\bibnamefont {Maris}}, \bibinfo {author} {\bibfnamefont {{\relax
  Ulf-G}.}~\bibnamefont {Mei\ss{}ner}}, \bibinfo {author} {\bibfnamefont
  {A.}~\bibnamefont {Nogga}}, \bibinfo {author} {\bibfnamefont
  {R.}~\bibnamefont {Roth}}, \bibinfo {author} {\bibfnamefont {R.}~\bibnamefont
  {Skibi\ifmmode~\acute{n}\else \'{n}\fi{}ski}}, \bibinfo {author}
  {\bibfnamefont {K.}~\bibnamefont {Topolnicki}}, \bibinfo {author}
  {\bibfnamefont {J.~P.}\ \bibnamefont {Vary}}, \bibinfo {author}
  {\bibfnamefont {K.}~\bibnamefont {Vobig}}, \ and\ \bibinfo {author}
  {\bibfnamefont {H.}~\bibnamefont {Wita\l{}a}} (\bibinfo {collaboration}
  {LENPIC Collaboration}),\ }\href {\doibase 10.1103/PhysRevC.98.014002}
  {\bibfield  {journal} {\bibinfo  {journal} {Phys. Rev. C}\ }\textbf {\bibinfo
  {volume} {98}},\ \bibinfo {pages} {014002} (\bibinfo {year} {2018})},\
  \Eprint {http://arxiv.org/abs/1802.08584} {arXiv:1802.08584 [nucl-th]}
  \BibitemShut {NoStop}%
\bibitem [{\citenamefont {Launey}\ \emph {et~al.}(2016)\citenamefont {Launey},
  \citenamefont {Dytrych},\ and\ \citenamefont {Draayer}}]{LauneyDD16}%
  \BibitemOpen
  \bibfield  {author} {\bibinfo {author} {\bibfnamefont {K.~D.}\ \bibnamefont
  {Launey}}, \bibinfo {author} {\bibfnamefont {T.}~\bibnamefont {Dytrych}}, \
  and\ \bibinfo {author} {\bibfnamefont {J.~P.}\ \bibnamefont {Draayer}},\
  }\href {\doibase 10.1016/j.ppnp.2016.02.001} {\bibfield  {journal} {\bibinfo
  {journal} {Prog. Part. Nucl. Phys.}\ }\textbf {\bibinfo {volume} {89}},\
  \bibinfo {pages} {101} (\bibinfo {year} {2016})},\ \Eprint
  {http://arxiv.org/abs/1612.04298} {arXiv:1612.04298 [nucl-th]} \BibitemShut
  {NoStop}%
\bibitem [{\citenamefont {Dytrych}\ \emph {et~al.}(2020)\citenamefont
  {Dytrych}, \citenamefont {Launey}, \citenamefont {Draayer}, \citenamefont
  {Rowe}, \citenamefont {Wood}, \citenamefont {Rosensteel}, \citenamefont
  {Bahri}, \citenamefont {Langr},\ and\ \citenamefont
  {Baker}}]{Dytrych:2020vkl}%
  \BibitemOpen
  \bibfield  {author} {\bibinfo {author} {\bibfnamefont {T.}~\bibnamefont
  {Dytrych}}, \bibinfo {author} {\bibfnamefont {K.~D.}\ \bibnamefont {Launey}},
  \bibinfo {author} {\bibfnamefont {J.~P.}\ \bibnamefont {Draayer}}, \bibinfo
  {author} {\bibfnamefont {D.~J.}\ \bibnamefont {Rowe}}, \bibinfo {author}
  {\bibfnamefont {J.~L.}\ \bibnamefont {Wood}}, \bibinfo {author}
  {\bibfnamefont {G.}~\bibnamefont {Rosensteel}}, \bibinfo {author}
  {\bibfnamefont {C.}~\bibnamefont {Bahri}}, \bibinfo {author} {\bibfnamefont
  {D.}~\bibnamefont {Langr}}, \ and\ \bibinfo {author} {\bibfnamefont {R.~B.}\
  \bibnamefont {Baker}},\ }\href {\doibase 10.1103/PhysRevLett.124.042501}
  {\bibfield  {journal} {\bibinfo  {journal} {Phys.\ Rev.\ Lett.}\ }\textbf
  {\bibinfo {volume} {124}},\ \bibinfo {pages} {042501} (\bibinfo {year}
  {2020})},\ \Eprint {http://arxiv.org/abs/1810.05757} {arXiv:1810.05757
  [nucl-th]} \BibitemShut {NoStop}%
\bibitem [{\citenamefont {Siciliano}\ and\ \citenamefont
  {Thaler}(1977)}]{Siciliano:1977zz}%
  \BibitemOpen
  \bibfield  {author} {\bibinfo {author} {\bibfnamefont {E.~R.}\ \bibnamefont
  {Siciliano}}\ and\ \bibinfo {author} {\bibfnamefont {R.~M.}\ \bibnamefont
  {Thaler}},\ }\href {\doibase 10.1103/PhysRevC.16.1322} {\bibfield  {journal}
  {\bibinfo  {journal} {Phys. Rev.}\ }\textbf {\bibinfo {volume} {C16}},\
  \bibinfo {pages} {1322} (\bibinfo {year} {1977})}\BibitemShut {NoStop}%
\bibitem [{\citenamefont {Burrows}\ \emph {et~al.}(2020)\citenamefont
  {Burrows}, \citenamefont {Baker}, \citenamefont {Elster}, \citenamefont
  {Weppner}, \citenamefont {Launey}, \citenamefont {Maris},\ and\ \citenamefont
  {Popa}}]{Burrows:2020qvu}%
  \BibitemOpen
  \bibfield  {author} {\bibinfo {author} {\bibfnamefont {M.}~\bibnamefont
  {Burrows}}, \bibinfo {author} {\bibfnamefont {R.~B.}\ \bibnamefont {Baker}},
  \bibinfo {author} {\bibfnamefont {{\relax Ch}.}~\bibnamefont {Elster}},
  \bibinfo {author} {\bibfnamefont {S.~P.}\ \bibnamefont {Weppner}}, \bibinfo
  {author} {\bibfnamefont {K.~D.}\ \bibnamefont {Launey}}, \bibinfo {author}
  {\bibfnamefont {P.}~\bibnamefont {Maris}}, \ and\ \bibinfo {author}
  {\bibfnamefont {G.}~\bibnamefont {Popa}},\ }\href {\doibase
  10.1103/PhysRevC.102.034606} {\bibfield  {journal} {\bibinfo  {journal}
  {Phys. Rev. C}\ }\textbf {\bibinfo {volume} {102}},\ \bibinfo {pages}
  {034606} (\bibinfo {year} {2020})},\ \Eprint
  {http://arxiv.org/abs/2005.00111} {arXiv:2005.00111 [nucl-th]} \BibitemShut
  {NoStop}%
\bibitem [{\citenamefont {Baker}\ \emph {et~al.}(2021)\citenamefont {Baker},
  \citenamefont {Burrows}, \citenamefont {Elster}, \citenamefont {Launey},
  \citenamefont {Maris}, \citenamefont {Popa},\ and\ \citenamefont
  {Weppner}}]{Baker:2021izp}%
  \BibitemOpen
  \bibfield  {author} {\bibinfo {author} {\bibfnamefont {R.~B.}\ \bibnamefont
  {Baker}}, \bibinfo {author} {\bibfnamefont {M.}~\bibnamefont {Burrows}},
  \bibinfo {author} {\bibfnamefont {{\relax Ch}.}~\bibnamefont {Elster}},
  \bibinfo {author} {\bibfnamefont {K.~D.}\ \bibnamefont {Launey}}, \bibinfo
  {author} {\bibfnamefont {P.}~\bibnamefont {Maris}}, \bibinfo {author}
  {\bibfnamefont {G.}~\bibnamefont {Popa}}, \ and\ \bibinfo {author}
  {\bibfnamefont {S.~P.}\ \bibnamefont {Weppner}},\ }\href {\doibase
  10.1103/PhysRevC.103.054314} {\bibfield  {journal} {\bibinfo  {journal}
  {Phys. Rev. C}\ }\textbf {\bibinfo {volume} {103}},\ \bibinfo {pages}
  {054314} (\bibinfo {year} {2021})},\ \Eprint
  {http://arxiv.org/abs/2102.01025} {arXiv:2102.01025 [nucl-th]} \BibitemShut
  {NoStop}%
\bibitem [{\citenamefont {Osborne}\ \emph {et~al.}(2004)\citenamefont
  {Osborne}, \citenamefont {Brady}, \citenamefont {Romero}, \citenamefont
  {Ullmann}, \citenamefont {Sorenson}, \citenamefont {Ling}, \citenamefont
  {King}, \citenamefont {Haight}, \citenamefont {Rapaport}, \citenamefont
  {Finlay}, \citenamefont {Bauge}, \citenamefont {Delaroche},\ and\
  \citenamefont {Koning}}]{Osborne:2004vd}%
  \BibitemOpen
  \bibfield  {author} {\bibinfo {author} {\bibfnamefont {J.~H.}\ \bibnamefont
  {Osborne}}, \bibinfo {author} {\bibfnamefont {F.~P.}\ \bibnamefont {Brady}},
  \bibinfo {author} {\bibfnamefont {J.~L.}\ \bibnamefont {Romero}}, \bibinfo
  {author} {\bibfnamefont {J.~L.}\ \bibnamefont {Ullmann}}, \bibinfo {author}
  {\bibfnamefont {D.~S.}\ \bibnamefont {Sorenson}}, \bibinfo {author}
  {\bibfnamefont {A.}~\bibnamefont {Ling}}, \bibinfo {author} {\bibfnamefont
  {N.~S.~P.}\ \bibnamefont {King}}, \bibinfo {author} {\bibfnamefont {R.~C.}\
  \bibnamefont {Haight}}, \bibinfo {author} {\bibfnamefont {J.}~\bibnamefont
  {Rapaport}}, \bibinfo {author} {\bibfnamefont {R.~W.}\ \bibnamefont
  {Finlay}}, \bibinfo {author} {\bibfnamefont {E.}~\bibnamefont {Bauge}},
  \bibinfo {author} {\bibfnamefont {J.~P.}\ \bibnamefont {Delaroche}}, \ and\
  \bibinfo {author} {\bibfnamefont {A.~J.}\ \bibnamefont {Koning}},\ }\href
  {\doibase 10.1103/PhysRevC.70.054613} {\bibfield  {journal} {\bibinfo
  {journal} {Phys. Rev. C}\ }\textbf {\bibinfo {volume} {70}},\ \bibinfo
  {pages} {054613} (\bibinfo {year} {2004})}\BibitemShut {NoStop}%
\bibitem [{\citenamefont {Furnstahl}\ \emph {et~al.}(2015)\citenamefont
  {Furnstahl}, \citenamefont {Klco}, \citenamefont {Phillips},\ and\
  \citenamefont {Wesolowski}}]{Furnstahl:2015rha}%
  \BibitemOpen
  \bibfield  {author} {\bibinfo {author} {\bibfnamefont {R.~J.}\ \bibnamefont
  {Furnstahl}}, \bibinfo {author} {\bibfnamefont {N.}~\bibnamefont {Klco}},
  \bibinfo {author} {\bibfnamefont {D.~R.}\ \bibnamefont {Phillips}}, \ and\
  \bibinfo {author} {\bibfnamefont {S.}~\bibnamefont {Wesolowski}},\ }\href
  {\doibase 10.1103/PhysRevC.92.024005} {\bibfield  {journal} {\bibinfo
  {journal} {Phys. Rev. C}\ }\textbf {\bibinfo {volume} {92}},\ \bibinfo
  {pages} {024005} (\bibinfo {year} {2015})},\ \Eprint
  {http://arxiv.org/abs/1506.01343} {arXiv:1506.01343 [nucl-th]} \BibitemShut
  {NoStop}%
\bibitem [{\citenamefont {Melendez}\ \emph {et~al.}(2017)\citenamefont
  {Melendez}, \citenamefont {Wesolowski},\ and\ \citenamefont
  {Furnstahl}}]{Melendez:2017phj}%
  \BibitemOpen
  \bibfield  {author} {\bibinfo {author} {\bibfnamefont {J.~A.}\ \bibnamefont
  {Melendez}}, \bibinfo {author} {\bibfnamefont {S.}~\bibnamefont
  {Wesolowski}}, \ and\ \bibinfo {author} {\bibfnamefont {R.~J.}\ \bibnamefont
  {Furnstahl}},\ }\href {\doibase 10.1103/PhysRevC.96.024003} {\bibfield
  {journal} {\bibinfo  {journal} {Phys. Rev. C}\ }\textbf {\bibinfo {volume}
  {96}},\ \bibinfo {pages} {024003} (\bibinfo {year} {2017})},\ \Eprint
  {http://arxiv.org/abs/1704.03308} {arXiv:1704.03308 [nucl-th]} \BibitemShut
  {NoStop}%
\bibitem [{\citenamefont {Melendez}\ \emph {et~al.}(2019)\citenamefont
  {Melendez}, \citenamefont {Furnstahl}, \citenamefont {Phillips},
  \citenamefont {Pratola},\ and\ \citenamefont
  {Wesolowski}}]{Melendez:2019izc}%
  \BibitemOpen
  \bibfield  {author} {\bibinfo {author} {\bibfnamefont {J.~A.}\ \bibnamefont
  {Melendez}}, \bibinfo {author} {\bibfnamefont {R.~J.}\ \bibnamefont
  {Furnstahl}}, \bibinfo {author} {\bibfnamefont {D.~R.}\ \bibnamefont
  {Phillips}}, \bibinfo {author} {\bibfnamefont {M.~T.}\ \bibnamefont
  {Pratola}}, \ and\ \bibinfo {author} {\bibfnamefont {S.}~\bibnamefont
  {Wesolowski}},\ }\href {\doibase 10.1103/PhysRevC.100.044001} {\bibfield
  {journal} {\bibinfo  {journal} {Phys. Rev. C}\ }\textbf {\bibinfo {volume}
  {100}},\ \bibinfo {pages} {044001} (\bibinfo {year} {2019})},\ \Eprint
  {http://arxiv.org/abs/1904.10581} {arXiv:1904.10581 [nucl-th]} \BibitemShut
  {NoStop}%
\bibitem [{\citenamefont {Epelbaum}\ \emph
  {et~al.}(2020{\natexlab{b}})\citenamefont {Epelbaum}, \citenamefont {Golak},
  \citenamefont {Hebeler}, \citenamefont {Kamada}, \citenamefont {Krebs},
  \citenamefont {Mei\ss{}ner}, \citenamefont {Nogga}, \citenamefont {Reinert},
  \citenamefont {Skibi\ifmmode~\acute{n}\else \'{n}\fi{}ski}, \citenamefont
  {Topolnicki}, \citenamefont {Volkotrub},\ and\ \citenamefont
  {Wita\l{}a}}]{Epelbaum:2019zqc}%
  \BibitemOpen
  \bibfield  {author} {\bibinfo {author} {\bibfnamefont {E.}~\bibnamefont
  {Epelbaum}}, \bibinfo {author} {\bibfnamefont {J.}~\bibnamefont {Golak}},
  \bibinfo {author} {\bibfnamefont {K.}~\bibnamefont {Hebeler}}, \bibinfo
  {author} {\bibfnamefont {H.}~\bibnamefont {Kamada}}, \bibinfo {author}
  {\bibfnamefont {H.}~\bibnamefont {Krebs}}, \bibinfo {author} {\bibfnamefont
  {{\relax Ulf-G}.}~\bibnamefont {Mei\ss{}ner}}, \bibinfo {author}
  {\bibfnamefont {A.}~\bibnamefont {Nogga}}, \bibinfo {author} {\bibfnamefont
  {P.}~\bibnamefont {Reinert}}, \bibinfo {author} {\bibfnamefont
  {R.}~\bibnamefont {Skibi\ifmmode~\acute{n}\else \'{n}\fi{}ski}}, \bibinfo
  {author} {\bibfnamefont {K.}~\bibnamefont {Topolnicki}}, \bibinfo {author}
  {\bibfnamefont {Y.}~\bibnamefont {Volkotrub}}, \ and\ \bibinfo {author}
  {\bibfnamefont {H.}~\bibnamefont {Wita\l{}a}},\ }\href {\doibase
  10.1140/epja/s10050-020-00102-2} {\bibfield  {journal} {\bibinfo  {journal}
  {Eur. Phys. J. A}\ }\textbf {\bibinfo {volume} {56}},\ \bibinfo {pages} {92}
  (\bibinfo {year} {2020}{\natexlab{b}})},\ \Eprint
  {http://arxiv.org/abs/1907.03608} {arXiv:1907.03608 [nucl-th]} \BibitemShut
  {NoStop}%
\bibitem [{\citenamefont {Maris}\ \emph {et~al.}(2021)\citenamefont {Maris},
  \citenamefont {Epelbaum}, \citenamefont {Furnstahl}, \citenamefont {Golak},
  \citenamefont {Hebeler}, \citenamefont {H\"{u}ther}, \citenamefont {Kamada},
  \citenamefont {Krebs}, \citenamefont {Mei\ss{}ner}, \citenamefont {Melendez},
  \citenamefont {Nogga}, \citenamefont {Reinert}, \citenamefont {Roth},
  \citenamefont {Skibi\ifmmode~\acute{n}\else \'{n}\fi{}ski}, \citenamefont
  {Soloviov}, \citenamefont {Topolnicki}, \citenamefont {Vary}, \citenamefont
  {Volkotrub}, \citenamefont {Wita\l{}a},\ and\ \citenamefont
  {Wolfgruber}}]{Maris:2020qne}%
  \BibitemOpen
  \bibfield  {author} {\bibinfo {author} {\bibfnamefont {P.}~\bibnamefont
  {Maris}}, \bibinfo {author} {\bibfnamefont {E.}~\bibnamefont {Epelbaum}},
  \bibinfo {author} {\bibfnamefont {R.~J.}\ \bibnamefont {Furnstahl}}, \bibinfo
  {author} {\bibfnamefont {J.}~\bibnamefont {Golak}}, \bibinfo {author}
  {\bibfnamefont {K.}~\bibnamefont {Hebeler}}, \bibinfo {author} {\bibfnamefont
  {T.}~\bibnamefont {H\"{u}ther}}, \bibinfo {author} {\bibfnamefont
  {H.}~\bibnamefont {Kamada}}, \bibinfo {author} {\bibfnamefont
  {H.}~\bibnamefont {Krebs}}, \bibinfo {author} {\bibfnamefont {{\relax
  Ulf-G}.}~\bibnamefont {Mei\ss{}ner}}, \bibinfo {author} {\bibfnamefont
  {J.~A.}\ \bibnamefont {Melendez}}, \bibinfo {author} {\bibfnamefont
  {A.}~\bibnamefont {Nogga}}, \bibinfo {author} {\bibfnamefont
  {P.}~\bibnamefont {Reinert}}, \bibinfo {author} {\bibfnamefont
  {R.}~\bibnamefont {Roth}}, \bibinfo {author} {\bibfnamefont {R.}~\bibnamefont
  {Skibi\ifmmode~\acute{n}\else \'{n}\fi{}ski}}, \bibinfo {author}
  {\bibfnamefont {V.}~\bibnamefont {Soloviov}}, \bibinfo {author}
  {\bibfnamefont {K.}~\bibnamefont {Topolnicki}}, \bibinfo {author}
  {\bibfnamefont {J.~P.}\ \bibnamefont {Vary}}, \bibinfo {author}
  {\bibfnamefont {{\relax Yu}.}~\bibnamefont {Volkotrub}}, \bibinfo {author}
  {\bibfnamefont {H.}~\bibnamefont {Wita\l{}a}}, \ and\ \bibinfo {author}
  {\bibfnamefont {T.}~\bibnamefont {Wolfgruber}},\ }\href {\doibase
  10.1103/PhysRevC.103.054001} {\bibfield  {journal} {\bibinfo  {journal}
  {Phys. Rev. C}\ }\textbf {\bibinfo {volume} {103}},\ \bibinfo {pages}
  {054001} (\bibinfo {year} {2021})},\ \Eprint
  {http://arxiv.org/abs/2012.12396} {arXiv:2012.12396 [nucl-th]} \BibitemShut
  {NoStop}%
\bibitem [{\citenamefont {Burrows}(2020)}]{BurrowsM:2020}%
  \BibitemOpen
  \bibfield  {author} {\bibinfo {author} {\bibfnamefont {M.}~\bibnamefont
  {Burrows}},\ }\emph {\bibinfo {title} {Ab Initio Leading Order Effective
  Interactions for Scattering of Nucleons from Light Nuclei}},\ \href@noop {}
  {Ph.D. thesis},\ \bibinfo  {school} {Ohio University} (\bibinfo {year}
  {2020})\BibitemShut {NoStop}%
\bibitem [{\citenamefont {Wolfenstein}(1954)}]{wolfenstein-1954}%
  \BibitemOpen
  \bibfield  {author} {\bibinfo {author} {\bibfnamefont {L.}~\bibnamefont
  {Wolfenstein}},\ }\href@noop {} {\bibfield  {journal} {\bibinfo  {journal}
  {Phys. Rev.}\ }\textbf {\bibinfo {volume} {96}},\ \bibinfo {pages} {1654}
  (\bibinfo {year} {1954})}\BibitemShut {NoStop}%
\bibitem [{\citenamefont {Chinn}\ \emph {et~al.}(1993)\citenamefont {Chinn},
  \citenamefont {Elster},\ and\ \citenamefont {Thaler}}]{Chinn:1993zza}%
  \BibitemOpen
  \bibfield  {author} {\bibinfo {author} {\bibfnamefont {C.~R.}\ \bibnamefont
  {Chinn}}, \bibinfo {author} {\bibfnamefont {{\relax Ch}.}~\bibnamefont
  {Elster}}, \ and\ \bibinfo {author} {\bibfnamefont {R.~M.}\ \bibnamefont
  {Thaler}},\ }\href {\doibase 10.1103/PhysRevC.47.2242} {\bibfield  {journal}
  {\bibinfo  {journal} {Phys. Rev. C}\ }\textbf {\bibinfo {volume} {47}},\
  \bibinfo {pages} {2242} (\bibinfo {year} {1993})}\BibitemShut {NoStop}%
\bibitem [{\citenamefont {Wolfenstein}\ and\ \citenamefont
  {Ashkin}(1952)}]{wolfenstein-ashkin}%
  \BibitemOpen
  \bibfield  {author} {\bibinfo {author} {\bibfnamefont {L.}~\bibnamefont
  {Wolfenstein}}\ and\ \bibinfo {author} {\bibfnamefont {J.}~\bibnamefont
  {Ashkin}},\ }\href@noop {} {\bibfield  {journal} {\bibinfo  {journal} {Phys.
  Rev.}\ }\textbf {\bibinfo {volume} {85}},\ \bibinfo {pages} {947} (\bibinfo
  {year} {1952})}\BibitemShut {NoStop}%
\bibitem [{\citenamefont {Launey}\ \emph {et~al.}(2014)\citenamefont {Launey},
  \citenamefont {Sarbadhicary}, \citenamefont {Dytrych},\ and\ \citenamefont
  {Draayer}}]{LauneySDD_CPC14}%
  \BibitemOpen
  \bibfield  {author} {\bibinfo {author} {\bibfnamefont {K.~D.}\ \bibnamefont
  {Launey}}, \bibinfo {author} {\bibfnamefont {S.}~\bibnamefont
  {Sarbadhicary}}, \bibinfo {author} {\bibfnamefont {T.}~\bibnamefont
  {Dytrych}}, \ and\ \bibinfo {author} {\bibfnamefont {J.~P.}\ \bibnamefont
  {Draayer}},\ }\href {\doibase https://doi.org/10.1016/j.cpc.2013.08.007}
  {\bibfield  {journal} {\bibinfo  {journal} {Comp. Phys. Comm.}\ }\textbf
  {\bibinfo {volume} {185}},\ \bibinfo {pages} {254 } (\bibinfo {year}
  {2014})}\BibitemShut {NoStop}%
\bibitem [{\citenamefont {Burrows}\ \emph {et~al.}(2019)\citenamefont
  {Burrows}, \citenamefont {Elster}, \citenamefont {Weppner}, \citenamefont
  {Launey}, \citenamefont {Maris}, \citenamefont {Nogga},\ and\ \citenamefont
  {Popa}}]{Burrows:2018ggt}%
  \BibitemOpen
  \bibfield  {author} {\bibinfo {author} {\bibfnamefont {M.}~\bibnamefont
  {Burrows}}, \bibinfo {author} {\bibfnamefont {{\relax Ch}.}~\bibnamefont
  {Elster}}, \bibinfo {author} {\bibfnamefont {S.~P.}\ \bibnamefont {Weppner}},
  \bibinfo {author} {\bibfnamefont {K.~D.}\ \bibnamefont {Launey}}, \bibinfo
  {author} {\bibfnamefont {P.}~\bibnamefont {Maris}}, \bibinfo {author}
  {\bibfnamefont {A.}~\bibnamefont {Nogga}}, \ and\ \bibinfo {author}
  {\bibfnamefont {G.}~\bibnamefont {Popa}},\ }\href@noop {} {\bibfield
  {journal} {\bibinfo  {journal} {Phys. Rev. C}\ }\textbf {\bibinfo {volume}
  {99}},\ \bibinfo {pages} {044603} (\bibinfo {year} {2019})},\ \Eprint
  {http://arxiv.org/abs/1810.06442} {arXiv:1810.06442 [nucl-th]} \BibitemShut
  {NoStop}%
\bibitem [{\citenamefont {Weppner}\ \emph {et~al.}(1998)\citenamefont
  {Weppner}, \citenamefont {Elster},\ and\ \citenamefont
  {Huber}}]{Weppner:1997wx}%
  \BibitemOpen
  \bibfield  {author} {\bibinfo {author} {\bibfnamefont {S.}~\bibnamefont
  {Weppner}}, \bibinfo {author} {\bibfnamefont {{\relax Ch}.}~\bibnamefont
  {Elster}}, \ and\ \bibinfo {author} {\bibfnamefont {D.}~\bibnamefont
  {Huber}},\ }\href {\doibase 10.1103/PhysRevC.57.1378} {\bibfield  {journal}
  {\bibinfo  {journal} {Phys. Rev. C}\ }\textbf {\bibinfo {volume} {57}},\
  \bibinfo {pages} {1378} (\bibinfo {year} {1998})},\ \Eprint
  {http://arxiv.org/abs/nucl-th/9712001} {arXiv:nucl-th/9712001} \BibitemShut
  {NoStop}%
\bibitem [{\citenamefont {Elster}\ \emph {et~al.}(1997)\citenamefont {Elster},
  \citenamefont {Weppner},\ and\ \citenamefont {Chinn}}]{Elster:1996xh}%
  \BibitemOpen
  \bibfield  {author} {\bibinfo {author} {\bibfnamefont {{\relax
  Ch}.}~\bibnamefont {Elster}}, \bibinfo {author} {\bibfnamefont {S.~P.}\
  \bibnamefont {Weppner}}, \ and\ \bibinfo {author} {\bibfnamefont {C.~R.}\
  \bibnamefont {Chinn}},\ }\href {\doibase 10.1103/PhysRevC.56.2080} {\bibfield
   {journal} {\bibinfo  {journal} {Phys. Rev. C}\ }\textbf {\bibinfo {volume}
  {56}},\ \bibinfo {pages} {2080} (\bibinfo {year} {1997})}\BibitemShut
  {NoStop}%
\bibitem [{\citenamefont {Finlay}\ \emph {et~al.}(1993)\citenamefont {Finlay},
  \citenamefont {Abfalterer}, \citenamefont {Fink}, \citenamefont {Montei},
  \citenamefont {Adami}, \citenamefont {Lisowski}, \citenamefont {Morgan},\
  and\ \citenamefont {Haight}}]{Finlay:1993hk}%
  \BibitemOpen
  \bibfield  {author} {\bibinfo {author} {\bibfnamefont {R.~W.}\ \bibnamefont
  {Finlay}}, \bibinfo {author} {\bibfnamefont {W.~P.}\ \bibnamefont
  {Abfalterer}}, \bibinfo {author} {\bibfnamefont {G.}~\bibnamefont {Fink}},
  \bibinfo {author} {\bibfnamefont {E.}~\bibnamefont {Montei}}, \bibinfo
  {author} {\bibfnamefont {T.}~\bibnamefont {Adami}}, \bibinfo {author}
  {\bibfnamefont {P.~W.}\ \bibnamefont {Lisowski}}, \bibinfo {author}
  {\bibfnamefont {G.~L.}\ \bibnamefont {Morgan}}, \ and\ \bibinfo {author}
  {\bibfnamefont {R.~C.}\ \bibnamefont {Haight}},\ }\href {\doibase
  10.1103/PhysRevC.47.237} {\bibfield  {journal} {\bibinfo  {journal} {Phys.
  Rev. C}\ }\textbf {\bibinfo {volume} {47}},\ \bibinfo {pages} {237} (\bibinfo
  {year} {1993})}\BibitemShut {NoStop}%
\bibitem [{\citenamefont {Pruitt}\ \emph {et~al.}(2020)\citenamefont {Pruitt},
  \citenamefont {Charity}, \citenamefont {Sobotka}, \citenamefont {Elson},
  \citenamefont {Hoff}, \citenamefont {Brown}, \citenamefont {Atkinson},
  \citenamefont {Dickhoff}, \citenamefont {Lee}, \citenamefont {Devlin},
  \citenamefont {Fotiades},\ and\ \citenamefont {Mosby}}]{Pruitt:2020kvc}%
  \BibitemOpen
  \bibfield  {author} {\bibinfo {author} {\bibfnamefont {C.~D.}\ \bibnamefont
  {Pruitt}}, \bibinfo {author} {\bibfnamefont {R.~J.}\ \bibnamefont {Charity}},
  \bibinfo {author} {\bibfnamefont {L.~G.}\ \bibnamefont {Sobotka}}, \bibinfo
  {author} {\bibfnamefont {J.~M.}\ \bibnamefont {Elson}}, \bibinfo {author}
  {\bibfnamefont {D.~E.~M.}\ \bibnamefont {Hoff}}, \bibinfo {author}
  {\bibfnamefont {K.~W.}\ \bibnamefont {Brown}}, \bibinfo {author}
  {\bibfnamefont {M.~C.}\ \bibnamefont {Atkinson}}, \bibinfo {author}
  {\bibfnamefont {W.~H.}\ \bibnamefont {Dickhoff}}, \bibinfo {author}
  {\bibfnamefont {H.~Y.}\ \bibnamefont {Lee}}, \bibinfo {author} {\bibfnamefont
  {M.}~\bibnamefont {Devlin}}, \bibinfo {author} {\bibfnamefont
  {N.}~\bibnamefont {Fotiades}}, \ and\ \bibinfo {author} {\bibfnamefont
  {S.}~\bibnamefont {Mosby}},\ }\href {\doibase 10.1103/PhysRevC.102.034601}
  {\bibfield  {journal} {\bibinfo  {journal} {Phys. Rev. C}\ }\textbf {\bibinfo
  {volume} {102}},\ \bibinfo {pages} {034601} (\bibinfo {year} {2020})},\
  \Eprint {http://arxiv.org/abs/2006.00024} {arXiv:2006.00024 [nucl-ex]}
  \BibitemShut {NoStop}%
\bibitem [{\citenamefont {Sakaguchi}\ \emph {et~al.}(1979)\citenamefont
  {Sakaguchi}, \citenamefont {Nakamura}, \citenamefont {Hatanaka},
  \citenamefont {Goto}, \citenamefont {Noro}, \citenamefont {Ohtani},
  \citenamefont {Sakamoto},\ and\ \citenamefont
  {Kobayashi}}]{Sakaguchi:1979fpk}%
  \BibitemOpen
  \bibfield  {author} {\bibinfo {author} {\bibfnamefont {H.}~\bibnamefont
  {Sakaguchi}}, \bibinfo {author} {\bibfnamefont {M.}~\bibnamefont {Nakamura}},
  \bibinfo {author} {\bibfnamefont {K.}~\bibnamefont {Hatanaka}}, \bibinfo
  {author} {\bibfnamefont {A.}~\bibnamefont {Goto}}, \bibinfo {author}
  {\bibfnamefont {T.}~\bibnamefont {Noro}}, \bibinfo {author} {\bibfnamefont
  {F.}~\bibnamefont {Ohtani}}, \bibinfo {author} {\bibfnamefont
  {H.}~\bibnamefont {Sakamoto}}, \ and\ \bibinfo {author} {\bibfnamefont
  {S.}~\bibnamefont {Kobayashi}},\ }\href {\doibase
  10.1016/0370-2693(79)90071-6} {\bibfield  {journal} {\bibinfo  {journal}
  {Phys. Lett. B}\ }\textbf {\bibinfo {volume} {89}},\ \bibinfo {pages} {40}
  (\bibinfo {year} {1979})}\BibitemShut {NoStop}%
\bibitem [{\citenamefont {Seifert}(1990)}]{Seifert:1990um}%
  \BibitemOpen
  \bibfield  {author} {\bibinfo {author} {\bibfnamefont {H.}~\bibnamefont
  {Seifert}},\ }\emph {\bibinfo {title} {Energy Dependence of the Effective
  Interaction for Nucleon-Nucleus Scattering}},\ \href@noop {} {Ph.D. thesis},\
  \bibinfo  {school} {University of Maryland} (\bibinfo {year}
  {1990})\BibitemShut {NoStop}%
\bibitem [{\citenamefont {Kelly}\ \emph {et~al.}(1989)\citenamefont {Kelly},
  \citenamefont {Bertozzi}, \citenamefont {Buti}, \citenamefont {Finn},
  \citenamefont {Hersman}, \citenamefont {Hyde-Wright}, \citenamefont {Hynes},
  \citenamefont {Kovash}, \citenamefont {Murdock}, \citenamefont {Norum},
  \citenamefont {Pugh}, \citenamefont {Rad}, \citenamefont {Bacher},
  \citenamefont {Emery}, \citenamefont {Foster}, \citenamefont {Jones},
  \citenamefont {Miller}, \citenamefont {Berman}, \citenamefont {Love},
  \citenamefont {Carr},\ and\ \citenamefont {Petrovich}}]{Kelly:1989zza}%
  \BibitemOpen
  \bibfield  {author} {\bibinfo {author} {\bibfnamefont {J.~J.}\ \bibnamefont
  {Kelly}}, \bibinfo {author} {\bibfnamefont {W.}~\bibnamefont {Bertozzi}},
  \bibinfo {author} {\bibfnamefont {T.~N.}\ \bibnamefont {Buti}}, \bibinfo
  {author} {\bibfnamefont {J.~M.}\ \bibnamefont {Finn}}, \bibinfo {author}
  {\bibfnamefont {F.~W.}\ \bibnamefont {Hersman}}, \bibinfo {author}
  {\bibfnamefont {C.}~\bibnamefont {Hyde-Wright}}, \bibinfo {author}
  {\bibfnamefont {M.~V.}\ \bibnamefont {Hynes}}, \bibinfo {author}
  {\bibfnamefont {M.~A.}\ \bibnamefont {Kovash}}, \bibinfo {author}
  {\bibfnamefont {B.}~\bibnamefont {Murdock}}, \bibinfo {author} {\bibfnamefont
  {B.~E.}\ \bibnamefont {Norum}}, \bibinfo {author} {\bibfnamefont
  {B.}~\bibnamefont {Pugh}}, \bibinfo {author} {\bibfnamefont {F.~N.}\
  \bibnamefont {Rad}}, \bibinfo {author} {\bibfnamefont {A.~D.}\ \bibnamefont
  {Bacher}}, \bibinfo {author} {\bibfnamefont {G.~T.}\ \bibnamefont {Emery}},
  \bibinfo {author} {\bibfnamefont {C.~C.}\ \bibnamefont {Foster}}, \bibinfo
  {author} {\bibfnamefont {W.~P.}\ \bibnamefont {Jones}}, \bibinfo {author}
  {\bibfnamefont {D.~W.}\ \bibnamefont {Miller}}, \bibinfo {author}
  {\bibfnamefont {B.~L.}\ \bibnamefont {Berman}}, \bibinfo {author}
  {\bibfnamefont {W.~G.}\ \bibnamefont {Love}}, \bibinfo {author}
  {\bibfnamefont {J.~A.}\ \bibnamefont {Carr}}, \ and\ \bibinfo {author}
  {\bibfnamefont {F.}~\bibnamefont {Petrovich}},\ }\href {\doibase
  10.1103/PhysRevC.39.1222} {\bibfield  {journal} {\bibinfo  {journal} {Phys.
  Rev. C}\ }\textbf {\bibinfo {volume} {39}},\ \bibinfo {pages} {1222}
  (\bibinfo {year} {1989})}\BibitemShut {NoStop}%
\bibitem [{\citenamefont {Kelly}\ \emph {et~al.}(1990)\citenamefont {Kelly},
  \citenamefont {Finn}, \citenamefont {Bertozzi}, \citenamefont {Buti},
  \citenamefont {Hersman}, \citenamefont {Hyde-Wright}, \citenamefont {Hynes},
  \citenamefont {Kovash}, \citenamefont {Murdock}, \citenamefont {Ulmer},
  \citenamefont {Bacher}, \citenamefont {Emery}, \citenamefont {Foster},
  \citenamefont {Jones}, \citenamefont {Miller},\ and\ \citenamefont
  {Berman}}]{Kelly:1990zza}%
  \BibitemOpen
  \bibfield  {author} {\bibinfo {author} {\bibfnamefont {J.~J.}\ \bibnamefont
  {Kelly}}, \bibinfo {author} {\bibfnamefont {J.~M.}\ \bibnamefont {Finn}},
  \bibinfo {author} {\bibfnamefont {W.}~\bibnamefont {Bertozzi}}, \bibinfo
  {author} {\bibfnamefont {T.~N.}\ \bibnamefont {Buti}}, \bibinfo {author}
  {\bibfnamefont {F.~W.}\ \bibnamefont {Hersman}}, \bibinfo {author}
  {\bibfnamefont {C.}~\bibnamefont {Hyde-Wright}}, \bibinfo {author}
  {\bibfnamefont {M.~V.}\ \bibnamefont {Hynes}}, \bibinfo {author}
  {\bibfnamefont {M.~A.}\ \bibnamefont {Kovash}}, \bibinfo {author}
  {\bibfnamefont {B.}~\bibnamefont {Murdock}}, \bibinfo {author} {\bibfnamefont
  {P.}~\bibnamefont {Ulmer}}, \bibinfo {author} {\bibfnamefont {A.~D.}\
  \bibnamefont {Bacher}}, \bibinfo {author} {\bibfnamefont {G.~T.}\
  \bibnamefont {Emery}}, \bibinfo {author} {\bibfnamefont {C.~C.}\ \bibnamefont
  {Foster}}, \bibinfo {author} {\bibfnamefont {W.~P.}\ \bibnamefont {Jones}},
  \bibinfo {author} {\bibfnamefont {D.~W.}\ \bibnamefont {Miller}}, \ and\
  \bibinfo {author} {\bibfnamefont {B.~L.}\ \bibnamefont {Berman}},\ }\href
  {\doibase 10.1103/PhysRevC.41.2504} {\bibfield  {journal} {\bibinfo
  {journal} {Phys. Rev. C}\ }\textbf {\bibinfo {volume} {41}},\ \bibinfo
  {pages} {2504} (\bibinfo {year} {1990})}\BibitemShut {NoStop}%
\bibitem [{\citenamefont {Ieiri}\ \emph {et~al.}(1987)\citenamefont {Ieiri},
  \citenamefont {Sakaguchi}, \citenamefont {Nakamura}, \citenamefont
  {Sakamoto}, \citenamefont {Ogawa}, \citenamefont {Yosol}, \citenamefont
  {Ichihara}, \citenamefont {Isshiki}, \citenamefont {Takeuchi}, \citenamefont
  {Togawa}, \citenamefont {Tsutsumi}, \citenamefont {Hirata}, \citenamefont
  {Nakano}, \citenamefont {Kobayashi}, \citenamefont {Noro},\ and\
  \citenamefont {Ikegami}}]{Ieiri1987253}%
  \BibitemOpen
  \bibfield  {author} {\bibinfo {author} {\bibfnamefont {M.}~\bibnamefont
  {Ieiri}}, \bibinfo {author} {\bibfnamefont {H.}~\bibnamefont {Sakaguchi}},
  \bibinfo {author} {\bibfnamefont {M.}~\bibnamefont {Nakamura}}, \bibinfo
  {author} {\bibfnamefont {H.}~\bibnamefont {Sakamoto}}, \bibinfo {author}
  {\bibfnamefont {H.}~\bibnamefont {Ogawa}}, \bibinfo {author} {\bibfnamefont
  {M.}~\bibnamefont {Yosol}}, \bibinfo {author} {\bibfnamefont
  {T.}~\bibnamefont {Ichihara}}, \bibinfo {author} {\bibfnamefont
  {N.}~\bibnamefont {Isshiki}}, \bibinfo {author} {\bibfnamefont
  {Y.}~\bibnamefont {Takeuchi}}, \bibinfo {author} {\bibfnamefont
  {H.}~\bibnamefont {Togawa}}, \bibinfo {author} {\bibfnamefont
  {T.}~\bibnamefont {Tsutsumi}}, \bibinfo {author} {\bibfnamefont
  {S.}~\bibnamefont {Hirata}}, \bibinfo {author} {\bibfnamefont
  {T.}~\bibnamefont {Nakano}}, \bibinfo {author} {\bibfnamefont
  {S.}~\bibnamefont {Kobayashi}}, \bibinfo {author} {\bibfnamefont
  {T.}~\bibnamefont {Noro}}, \ and\ \bibinfo {author} {\bibfnamefont
  {H.}~\bibnamefont {Ikegami}},\ }\href {\doibase
  https://doi.org/10.1016/0168-9002(87)90744-3} {\bibfield  {journal} {\bibinfo
   {journal} {Nucl. Instrum. Methods Phys. Res. A}\ }\textbf {\bibinfo {volume}
  {257}},\ \bibinfo {pages} {253} (\bibinfo {year} {1987})}\BibitemShut
  {NoStop}%
\bibitem [{\citenamefont {Meyer}\ \emph {et~al.}(1983)\citenamefont {Meyer},
  \citenamefont {Schwandt}, \citenamefont {Jacobs},\ and\ \citenamefont
  {Hall}}]{Meyer:1983kd}%
  \BibitemOpen
  \bibfield  {author} {\bibinfo {author} {\bibfnamefont {H.~O.}\ \bibnamefont
  {Meyer}}, \bibinfo {author} {\bibfnamefont {P.}~\bibnamefont {Schwandt}},
  \bibinfo {author} {\bibfnamefont {W.~W.}\ \bibnamefont {Jacobs}}, \ and\
  \bibinfo {author} {\bibfnamefont {J.~R.}\ \bibnamefont {Hall}},\ }\href
  {\doibase 10.1103/PhysRevC.27.459} {\bibfield  {journal} {\bibinfo  {journal}
  {Phys. Rev. C}\ }\textbf {\bibinfo {volume} {27}},\ \bibinfo {pages} {459}
  (\bibinfo {year} {1983})}\BibitemShut {NoStop}%
\bibitem [{\citenamefont {Sakaguchi}\ \emph {et~al.}(1986)\citenamefont
  {Sakaguchi}, \citenamefont {Yosoi}, \citenamefont {Nakamura}, \citenamefont
  {Noro}, \citenamefont {Sakamoto}, \citenamefont {Ichihara}, \citenamefont
  {Ieiri}, \citenamefont {Takeuchi}, \citenamefont {Togawa}, \citenamefont
  {T.}, \citenamefont {Ikegami},\ and\ \citenamefont
  {Kobayashi}}]{Sakaguchi:1986}%
  \BibitemOpen
  \bibfield  {author} {\bibinfo {author} {\bibfnamefont {H.}~\bibnamefont
  {Sakaguchi}}, \bibinfo {author} {\bibfnamefont {M.}~\bibnamefont {Yosoi}},
  \bibinfo {author} {\bibfnamefont {M.}~\bibnamefont {Nakamura}}, \bibinfo
  {author} {\bibfnamefont {T.}~\bibnamefont {Noro}}, \bibinfo {author}
  {\bibfnamefont {H.}~\bibnamefont {Sakamoto}}, \bibinfo {author}
  {\bibfnamefont {T.}~\bibnamefont {Ichihara}}, \bibinfo {author}
  {\bibfnamefont {M.}~\bibnamefont {Ieiri}}, \bibinfo {author} {\bibfnamefont
  {Y.}~\bibnamefont {Takeuchi}}, \bibinfo {author} {\bibfnamefont
  {H.}~\bibnamefont {Togawa}}, \bibinfo {author} {\bibfnamefont
  {T.}~\bibnamefont {T.}}, \bibinfo {author} {\bibfnamefont {H.}~\bibnamefont
  {Ikegami}}, \ and\ \bibinfo {author} {\bibfnamefont {S.}~\bibnamefont
  {Kobayashi}},\ }\href@noop {} {\bibfield  {journal} {\bibinfo  {journal} {J.
  Phys. Soc. Japan Suppl.}\ }\textbf {\bibinfo {volume} {55}},\ \bibinfo
  {pages} {61} (\bibinfo {year} {1986})}\BibitemShut {NoStop}%
\bibitem [{\citenamefont {Machleidt}(2001)}]{Machleidt:2000ge}%
  \BibitemOpen
  \bibfield  {author} {\bibinfo {author} {\bibfnamefont {R.}~\bibnamefont
  {Machleidt}},\ }\href {\doibase 10.1103/PhysRevC.63.024001} {\bibfield
  {journal} {\bibinfo  {journal} {Phys. Rev. C}\ }\textbf {\bibinfo {volume}
  {63}},\ \bibinfo {pages} {024001} (\bibinfo {year} {2001})},\ \Eprint
  {http://arxiv.org/abs/nucl-th/0006014} {arXiv:nucl-th/0006014 [nucl-th]}
  \BibitemShut {NoStop}%
\end{thebibliography}%
